\documentclass[12pt,natbib]{emulateapj}

\def \HST{{\emph{HST}}}
\def \Spitzer{{\emph{Spitzer}}}
\def \tf {{24 $\mu$m}}
\def \um {{$\mu$m}}
\def \boot {{Bo\"otes}}

\begin{document}
\slugcomment{03/13/09}

\title{High Redshift Dust Obscured Galaxies, A Morphology-SED Connection Revealed by Keck Adaptive Optics \footnote{Some of the data presented herein were obtained at the W.M. Keck Observatory, which is operated as a scientific partnership among the California Institute of Technology, the University of California and the National Aeronautics and Space Administration. The Observatory was made possible by the generous financial support of the W.M. Keck Foundation.}}
 
\author{J. Melbourne \altaffilmark{1}, S. Bussman \altaffilmark{2}, K. Brand \altaffilmark {3}, V. Desai \altaffilmark{4}, L. Armus \altaffilmark{4}, Arjun Dey \altaffilmark{5},  B. T. Jannuzi \altaffilmark{5}, J. R. Houck \altaffilmark{6}, K. Matthews\altaffilmark{1}, B. T. Soifer \altaffilmark{1,4}}

\altaffiltext{1}{Caltech Optical Observatories, Division of Physics, Mathematics and Astronomy, Mail Stop 320-47, California Institute of Technology, Pasadena, CA 91125, jmel@caltech.edu,  bts@submm.caltech.edu, kym@caltech.edu}

\altaffiltext{2}{University of Arizona, LBT Observatory, 933 N. Cherry Ave. Tuscon, AZ 85721-0065, dthompson@as.arizona.edu}

\altaffiltext{3}{Space Telescope Science Institute, Baltimore, MD 21218, brand@stsci.edu}

\altaffiltext{4}{Spitzer Science Center, Mail Stop 314-6, California Institute of Technology, Pasadena, CA 91125, vandesai@gmail.com, lee@ipac.caltech.edu, bts@ipac.caltech.edu}

\altaffiltext{5}{National Optical Astronomy Observatory, P.O. Box 26732, Tucson, AZ 85726-6732, dey@noao.edu, jannuzi@noao.edu}

\altaffiltext{6}{Astronomy Department, Cornell University, Ithaca, NY 14853, jrh13@cornell.edu}

\begin{abstract}
A simple optical to mid-IR color selection, $R-[24] > 14$, i.e. $f_{\nu}$(\tf)$ / f_{\nu} (R) \ga 1000$, identifies highly dust obscured galaxies (DOGs) with typical redshifts of $z\sim2\pm0.5$.   Extreme mid-IR luminosities ($L_{IR} > 10^{12-14}$) suggest that DOGs are powered by a combination of AGN and star formation, possibly driven by mergers.  In an effort to compare their photometric properties with their rest frame optical morphologies, we obtained high spatial resolution ($0.05 -0.1 \arcsec$) Keck Adaptive Optics (AO) $K'$-band images of 15 DOGs.  The images reveal a wide range of morphologies, including: small exponential disks (8 of 15), small ellipticals (4 of 15), and unresolved sources (2 of 15).  One particularly diffuse source could not be classified because of low signal to noise ratio. We find a statistically significant correlation between galaxy concentration and mid-IR luminosity, with the most luminous DOGs  exhibiting higher concentration and smaller physical size.  DOGs with high concentration also tend to have spectral energy distributions (SEDs) suggestive of AGN activity.  Thus central AGN light may be biasing the morphologies of the more luminous DOGs to higher concentration.  Conversely, more diffuse DOGs tend to show an SED shape suggestive of star formation.  Two of fifteen in the sample show multiple resolved components with separations of $\sim1$ kpc,   circumstantial evidence for ongoing mergers.  
\end{abstract}


\section{Introduction}
Recent \Spitzer\ Space Telescope \tf\ images of extragalactic survey fields have revealed extremely dust obscured galaxies \citep[DOGs; e.g.][]{Houck05,Dey08, Fiore08}.  Defined by very red optical to infrared (IR) colors, $R-[24] > 14$, i.e. $f_{\nu}$(\tf)$ / f_{\nu} (R) \ga 1000$, DOGs are redder than the typical low redshift ultra-luminous infrared galaxy \citep[ULIRG; $L_{IR} =10^{12-13}$;][]{Dey08}.  The  $\sim9$ square degree \boot\ field of the NOAO Deep Wide-Field Survey \citep[NDWFS;][]{JannuziDey99} contains $\sim 2600$ DOGs (out  of $\sim20000$ \tf\ sources with a limiting flux density of 0.3 mJy).  Spectroscopic redshifts of 86 DOGs in \boot\ reveal a $< z > \ =1.99$ with $\sigma_z =0.5$ \citep{Dey08}.  At these distances, the implied total IR luminosities of the DOGs are $L_{IR} > 10^{12-14}$, typically in excess of low redshift ULIRGS.  The extreme luminosities and colors suggest rapid AGN accretion and/or intense star-formation activity, heavily obscured by dust at rest-frame optical and UV wavelengths. 

Optical-IR spectral-energy distributions (SEDs) of the DOGs suggest two classes.    So called ``power-law'' sources show a continuous rise in flux density to longer wavelengths.   \Spitzer\ IRS mid-IR spectra of power-law DOGs show strong silicate absorption, an indicator of AGN activity \citep{Houck05,Weedman06}.  In contrast, so called ``bump'' sources exhibit a rest-frame 1.6 \um\ flux excess thought to be produced by the stellar photospheres of cooler stars.  \Spitzer\ IRS spectra of bump sources show strong PAH features suggestive of rapid star formation \citep[e.g.][; Desai et al. 2009, in preparation]{Farrah08}.  At these redshifts,  the ``bump'' appears in the \Spitzer\ IRAC bands ($3.5- 8.0$ \um).  Based on: (1) space densities (Dey et al. 2008); (2) clustering strength (Brodwin et al. 2008); and (3) mid/far IR SEDs (Pope et al. 2008), DOGS, may be linked to the  $z=2$ sub-mm galaxies (SMGs), which are thought to be experiencing merger driven star formation rates of up to several thousand solar masses per year. As with ULIRGs at low redshift \citep{Armus07}, most DOGs are likely to have some combination of AGN and star formation activity.  For example, a study of the average (stacked) x-ray properties of DOGs by \citet{Fiore08}  showed that even lower luminosity DOGs exhibit hard x-ray sources, suggestive of AGN activity. \citet{Pope08}, however, suggests that fainter DOGs are primarily powered by star formation even at x-ray wavelengths.  

While significant progress has been made in determining the bulk properties of the DOGs, less is known about the specific triggers of the AGN and star formation.  High spatial resolution imaging may help identify potential triggers.  For instance, by analogy with local ULIRGs \citep{Sanders88} and as might be expected from hierarchical structure formation models \citep[e.g.][]{Somerville01},  DOGs may be the product of gas rich mergers.   High resolution imaging may reveal multi-component systems and tidal features typical of mergers.  Alternatively DOGs may be the product of the monolithic collapse of primordial clouds seen at the time of first assembly \citep[e.g.][]{Eggen62}.  If the DOGs represent the formation of today's most massive systems (typically elliptical galaxies), they may already preferentially exhibit elliptical morphologies at $z=2$ \citep{Zirm07}.    

\citet[][hereafter Mel08]{Melbourne08b} and Bussman et al. (2009, in preparation) describe our early efforts to obtain high spatial resolution Keck Adaptive Optics, ($K'$-band) and Hubble Space Telescope (\HST; NICMOS $H$-band; ACS $i$-band; and WFPC2 $V$-band) images of the DOGs.   These studies focussed on power-law sources and, especially at longer wavelengths ($H$ and $K'$-bands), typically found smooth systems with small sizes (i.e. typically $R_{1/2}<3$ kpc). The \HST\ optical data (rest UV) tended to contain more substructures as might be expected from multiple star forming regions.   

Unfortunately, there were significant limitations to these first efforts.  The Keck Adaptive Optics (AO) imaging in particular was a very small sample size, containing only three DOGs.  The AO sample  were among the brightest in the entire NDWFS sample, none with a known redshift.  While the sample size for the \HST\ study was significantly larger, 30 systems, the \HST\ data also had limitations.  The very faint optical (rest UV) fluxes for the DOGs resulted in low signal-to-noise ratio (S/N) in the optical \HST\ imaging, especially for the WFPC2 data.  The \HST\ NICMOS ($H$-band, rest blue) images of the DOGs were higher S/N than the optical data, but the NICMOS $H$-band imaging had significantly lower spatial resolution ($\sim0.15\arcsec$) than the \HST\ optical or Keck AO $K$-band ($\sim0.05\arcsec$) data.

This follow-up paper presents a significantly larger sample of DOGs with Keck AO imaging in the $K$-band, now 15, including 2 with overlap with the \HST\ sample.  The sample now spans the range of \tf\ flux densities  for NDWFS DOGs ( 0.3 mJy $<  f_{\nu}$(\tf) $< 5$ mJy), and includes 6 potential bump sources.  Section 3  gives the Keck AO derived morphologies of these systems and correlates them with other observed properties of the sample.  Results are discussed in Section 4.

Throughout, we report Vega magnitudes and assume a $\Lambda$ cold-dark-matter cosmology: a flat universe with $H_0= 70$ km/s/Mpc, $\Omega_m=0.3$, and $\Omega_{\Lambda}=0.7$.

\begin{figure}
\centering
\includegraphics[scale=0.5]{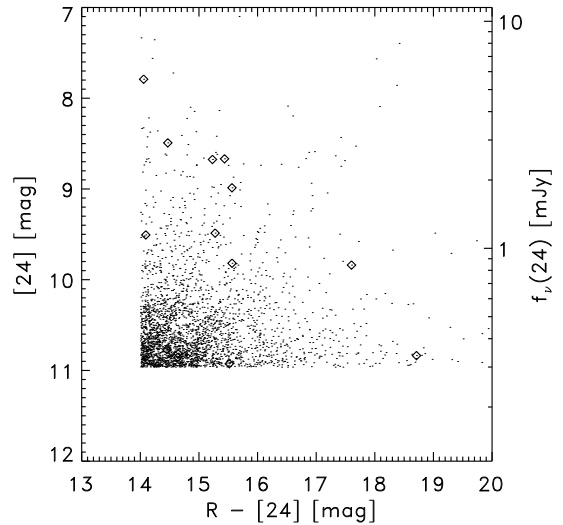}
\caption{\label{fig:cmd} Optical/IR color-magnitude diagram for DOGs in \boot\ (points).  The DOGs in the AO sample are shown as diamonds.  The data show both the sharp \tf\ flux limit and the color definition for DOGs.  The AO sample spans the full range of DOGs in color-magnitude space.}
\end{figure}

\section{The Data}

\subsection{Sample Selection}
Adaptive optics makes use of a reference source to track and correct for atmospheric turbulence.  AO corrected images approach the  diffraction limited resolution of the telescope which for Keck in the $K$-band is $\sim0.05\arcsec$, similar the the resolution of \HST\ in the optical.  Natural Guide Star (NGS) facilities use a star near the science target as a reference source.  NGS guide stars need to be fairly bright \citep[typically $R<14.5$ for Keck,][]{Wizinowich00,vanDam04} for the AO system to work.  In addition, the stars need to be near (typically within $30\arcsec$) the science target because the AO correction falls off with distance from the guide star (anisoplanetism).   A Laser Guide Star (LGS) facility uses a laser to produce a reference source high in the Earth's atmosphere for the AO correction.  The laser spot is used to track the high order wavefront errors produced by the turbulent atmosphere.  LGS systems still require a ``tip-tilt'' reference star to correct the low order (tip/tilt) terms  and eliminate image motion.  However these tip-tilt stars can be fainter and significantly further from the science target than NGS guide stars \citep[$R<17.5$ and $d<55\arcsec$ for Keck,][]{Wizinowich06}.  Another benefit of the LGS facility is that the laser spot is at the location of the science target, reducing the affects of anisoplanetism.  Of the $\sim2600$ DOGs in the \boot\ field of the NDWFS \citep{JannuziDey99}, roughly a third (918) are observable with with Keck LGS AO (using tip-tilt reference stars with $R < 17.5$ [mag], at an angular separation $d<55 \arcsec$).  Only 15 are observable with Keck NGS AO (using NGS guide stars with $R < 14.5$ [mag], at an angular separation $d<30 \arcsec$).  

\begin{figure*}
%\plotfiddle{DOG_01_d_contour.ps}{0.1in}{0}{200}{100}{0}{0}
%\plotfiddle{DOG_01_d_profile_.ps}{1in}{0}{50}{50}{350}{-150}
\includegraphics[scale=0.8]{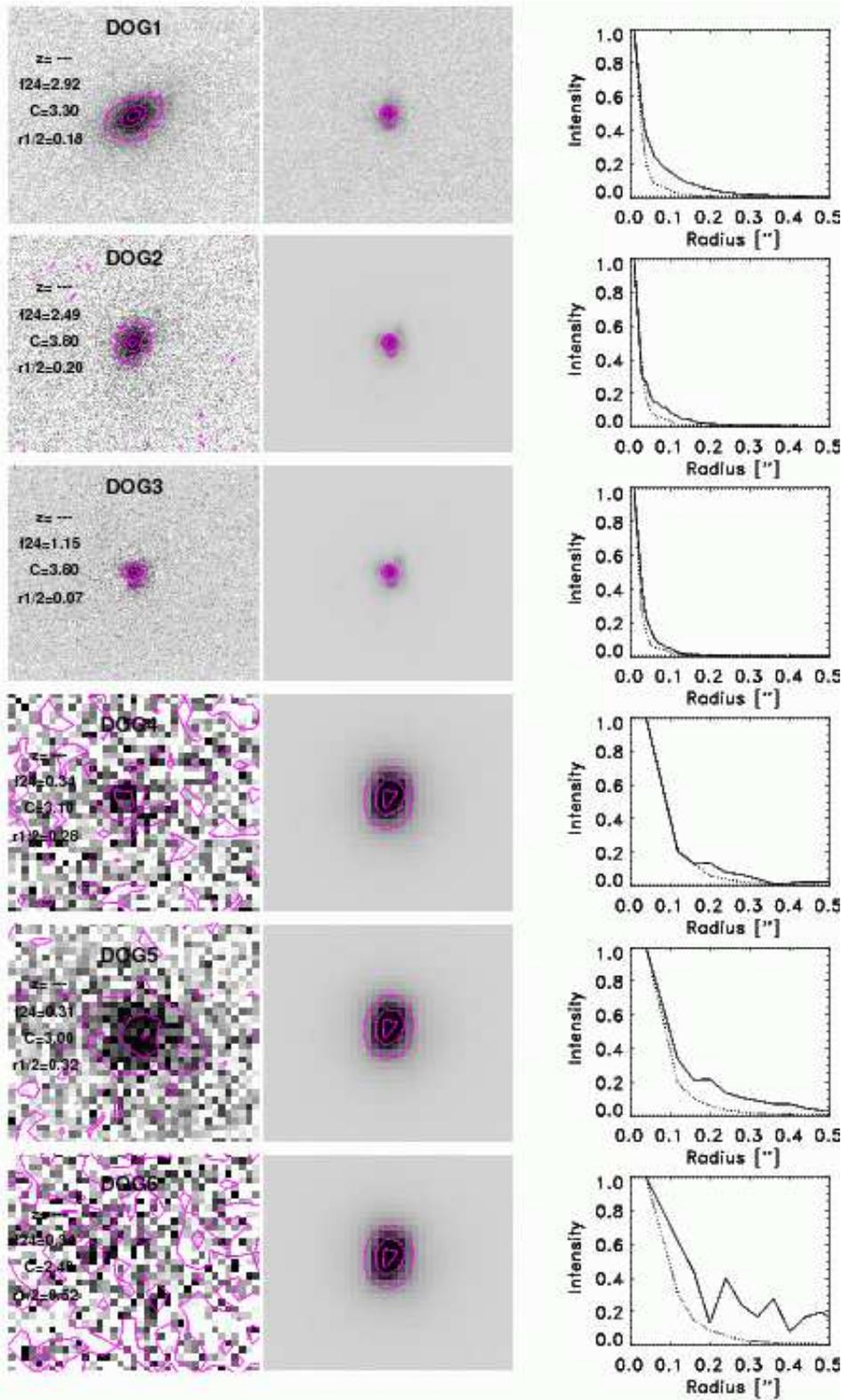}

\caption{\label{fig:im} The 2007 Keck AO observations of 6 DOGs (left) and their associated PSFs (middle).  Contours are over-plotted to demonstrate the differences between the galaxy and PSF.  1D radial intensity profiles of the DOGs (solid line) and PSFs (dotted line) are shown (right).  The DOGs tend to have more extended profiles than their associated PSFs.  The box size is $\sim1.5\arcsec$ on a side.  DOGs 1 - 3 were observed with the narrow-field camera (pixscale $=0.01\arcsec$ pix$^{-1}$) in LGS mode, while DOGs 4 - 6 were observed with the wide-field camera (pixscale $=0.04\arcsec$ pix$^{-1}$) in NGS mode.  } 
\end{figure*}

\begin{figure*}
\includegraphics[scale=0.8]{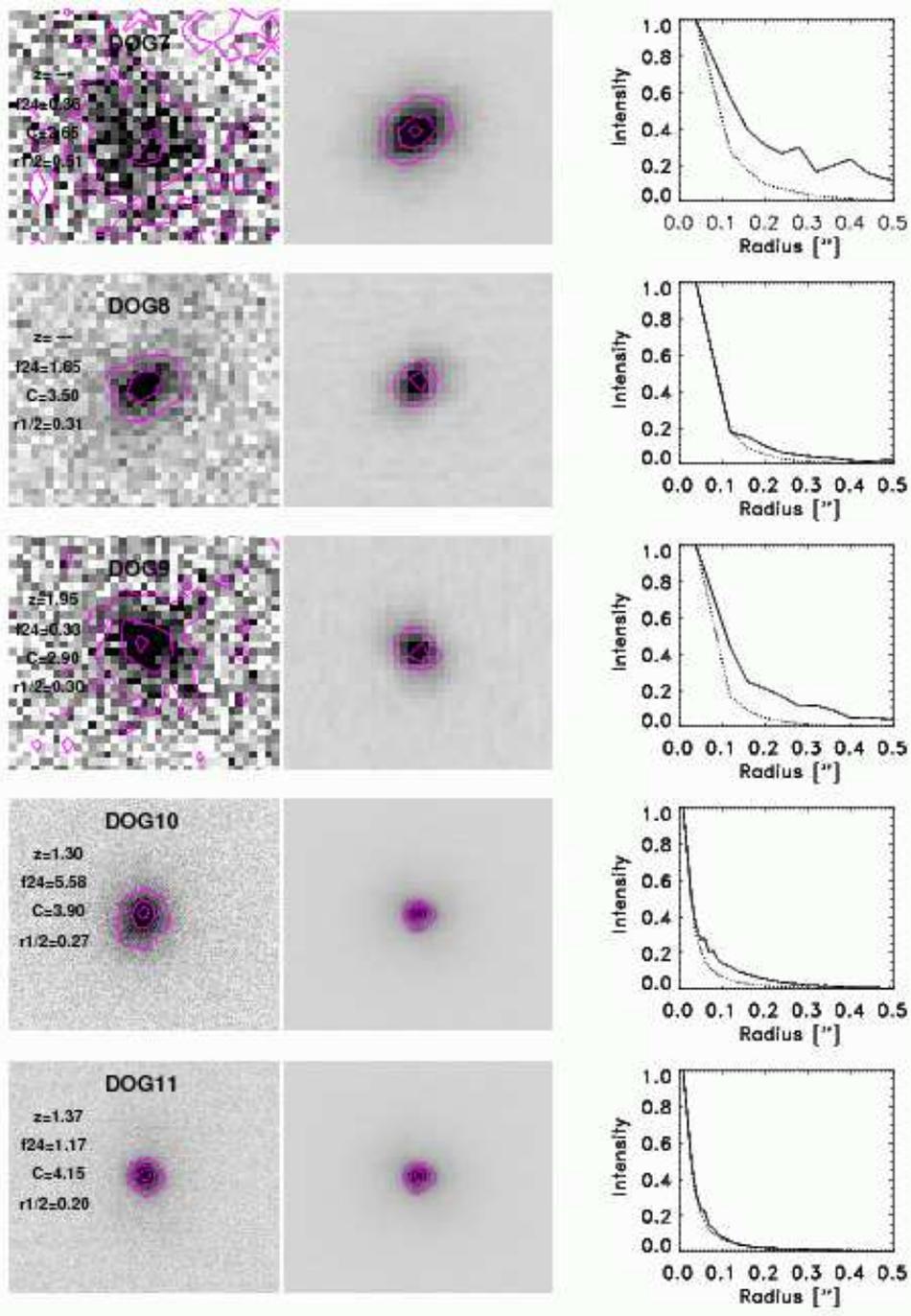}
\caption{\label{fig:im2} Same as Figure \ref{fig:im} only now for 2008 observations.  DOGs 7-9 were observed with the wide field camera (pixscale $=0.04\arcsec$ pix$^{-1}$) in NGS mode, while DOGs 10 and 11 were observed in the narrow-field camera (pixscale $=0.01\arcsec$ pix$^{-1}$) in LGS mode.
}
\end{figure*}

\begin{figure*}
\includegraphics[scale=0.8]{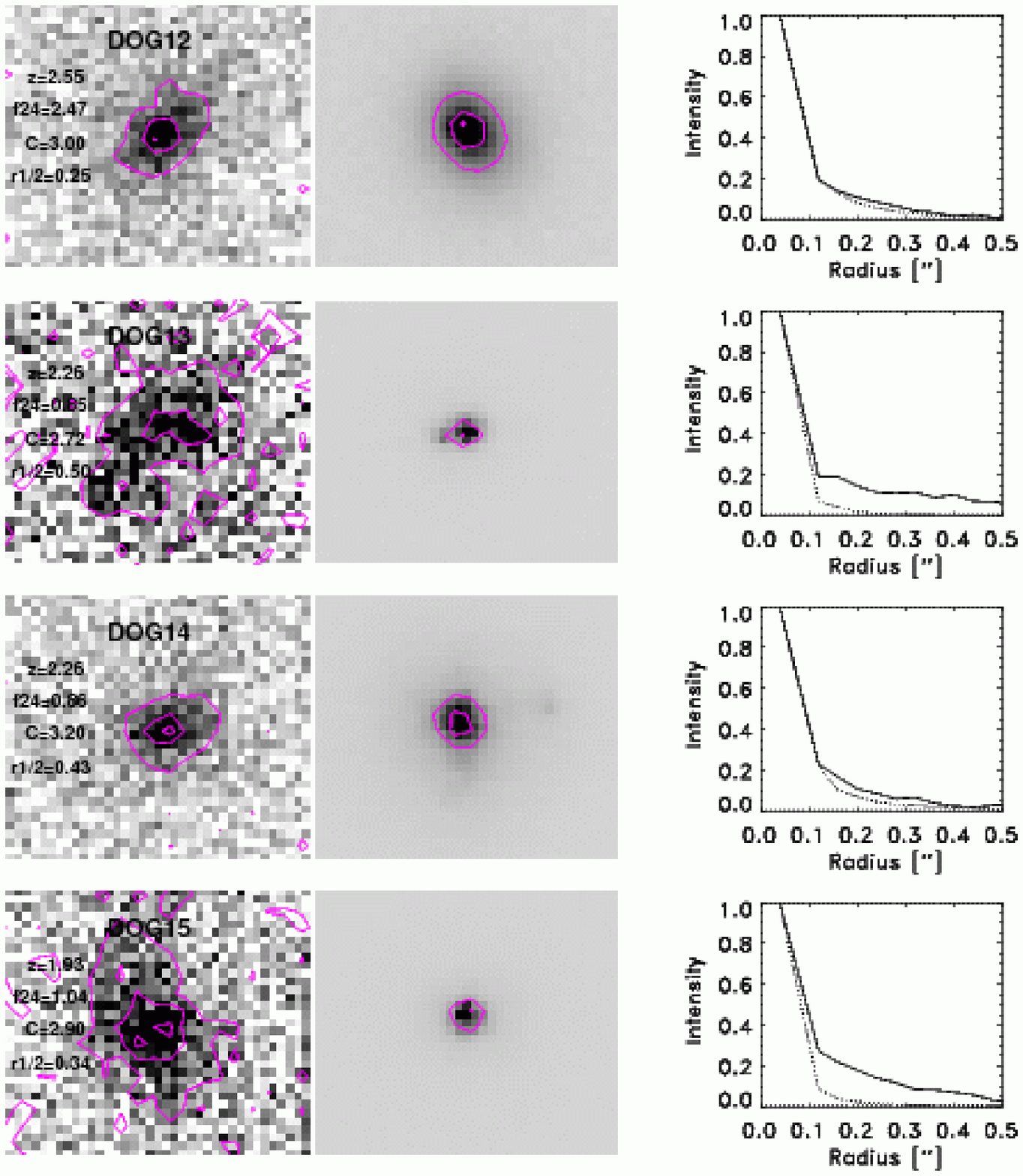}
\caption{\label{fig:im3} Additional 2008 observations made with the wide field camera (pixscale $=0.04\arcsec$ pix$^{-1}$).  DOGs 12 - 14 were made in the LGS mode, while DOG 15 was made in the NGS mode.
}
\end{figure*}

 A wide variety of selection criteria were used to select the DOGs in this paper.  Fourteen were chosen from the \boot\ field of the NDWFS \citep{JannuziDey99}.  One was chosen from the Spitzer First Look Survey \citep[FLS;][]{Yan07}.   In 2007, the three LGS targets were selected to provide the best possible S/N and AO performance, i.e. among the brightest DOGs in the $K$-band,  and close to a bright AO tip-tilt star. Initial results on these sources were given in \citet{Melbourne08b}. Three NGS targets were selected to be near to bright NGS guide stars, with no selection for $K$-band brightness.  None of the DOGs observed in 2007 had spectroscopic redshifts.  

The 2008 AO selection criteria were relaxed to primarily select DOGs with existing spectroscopic redshifts, including two NGS targets and five LGS targets \citep{Yan07,Dey08}.  Two additional NGS targets observed in 2008  did not have spectroscopic redshifts.  Complicating the 2008 selection criteria, two DOGs were selected because of interesting morphology in existing optical and near-IR \HST\ imaging.

While the selection criteria are non-uniform, the final sample of 15 DOGs spans the full range of \tf\ fluxes and $R - [24]$ colors of the larger \boot\ DOG sample (Figure \ref{fig:cmd}).

\subsection{AO Observations}
Keck AO observations were obtained for 15 DOGs over the course of 5 nights in the Spring of 2007 and 2008.  Table \ref{tab:obs} details the observing conditions on these dates.     Because the first two nights in the 2007 campaign had heavy extinction by clouds (over 1 magnitude of extinction in $R$), we could not propagate the laser.  Instead we observed 3 NGS targets.  Conditions were better during the final 2007 observing night, which produced images of 3 DOGs from the LGS target list.  In 2008, we observed 4 DOGS with the NGS facility and 5 with the LGS facility, all under good observing conditions.   For a more detailed description of our observing strategies with the Keck LGS AO facility see \citet{Melbourne08a, Melbourne08b}.    
 \begin{deluxetable}{clll}
\tabletypesize{\small}
\tablecaption{Keck Observation Run Summary \label{tab:obs}}
\tablehead{\colhead{Obsrun} & \colhead{UT Date} & \colhead{Run Type}  & \colhead{Conditions} }
\startdata
1 & May 12, 2007 & NGS & thick cirrus \\
2 & May 21, 2007 & LGS & scattered clouds \\
3 & May 19, 2008 & NGS & clear \\
4 & May 20, 2008 & LGS & clear \\
5 & June 2, 2008 & LGS & clear \\
\enddata			 
\end{deluxetable}

 \begin{deluxetable*}{c|c|ccc|cccc|c|cc}
\tabletypesize{\tiny}
\tablecaption{Observing Summary \label{tab:obs2}}
\tablehead{ & & &  & & & & & \colhead{per pix} &\multicolumn{1}{c}{PSF} & \multicolumn{2}{c}{TipTilt or NGS} \\
\colhead{object} & \colhead{Official Name \tablenotemark{a}} & \colhead{z} & \colhead{f24} &\colhead{$R-[24]$}&\colhead{Obsrun}  & \colhead{exptime } & \colhead{pixscale} & \colhead{SB limit} & \colhead{FWHM} &  \colhead{$R$} & \colhead{sep} \\ 
& & & \colhead{[mJ]}& [mag] & & \colhead{[min]} & \colhead{[$\arcsec$/pix]} & \colhead{[mag/($\arcsec)^2$]}&\colhead{ [$\arcsec$]} 
 &\colhead{ [mag]} &\colhead{[$\arcsec$]}}

%\tablehead{\colhead{object} & \colhead{Official Name \tablenotemark{a}} & \colhead{Obsrun}  & \colhead{Exptime}  }
\startdata
DOG 1 &	 SST24 J143234.9+333637 &    -   & 2.924 & 14.47 & 2 (LGS)& 30 & 0.01& 17.4& 0.057 &  15.0 & 26\\
DOG 2 &  SST24 J142801.0+341525 &    -   & 2.487 & 15.44 & 2 (LGS)& 30 & 0.01& 17.4& 0.053 &  14.3 & 20\\
DOG 3 &  SST24 J142944.9+324332 &    -   & 1.148 & 14.09 & 2 (LGS)& 9  & 0.01& 16.9& 0.053 & 14.5 & 27\\
DOG 4 &  SST24 J143117.1+332024 &    -   & 0.344 & 14.64 & 1 (NGS)& 60 & 0.04& 18.5& 0.088 &  11.6 & 28\\
DOG 5 &  SST24 J142825.5+343830 &    -   & 0.310 & 15.52 & 1 (NGS)& 68 & 0.04& 19.3& 0.088 &  13.7 & 22\\
DOG 6 &  SST24 J142616.2+352116 &    -   & 0.337 & 14.30 & 1 (NGS)& 60 & 0.04& 19.0& 0.088 &  13.0 & 29\\
DOG 7 &  SST24 J142925.9+345151 &    -   & 0.359 & 14.45 & 3 (NGS)& 96 & 0.04& 19.4& 0.104 &     14.2 & 23\\
DOG 8 &	 SST24 J143032.8+340046	&    -   & 1.854 & 15.56 & 3 (NGS)& 81 & 0.04& 19.1& 0.072     & 14.0 & 23\\
DOG 9 &	 SST24 J143641.1+350207	& 1.948  & 0.332 & 14.64 & 3 (NGS)& 63 & 0.04& 19.2& 0.072     & 14.2 & 24\\
DOG 10&	 SST24 J143335.6+354243	& 1.297  & 5.577 & 14.06 & 4 (LGS)& 24 & 0.01& 17.2& 0.052     & 15.5 & 47\\
DOG 11&	 SST24 J143027.2+344008	& 1.370  & 1.169 & 15.28 & 4 (LGS)& 39 & 0.01& 17.4& 0.051     & 16.8 & 41\\
DOG 12&	 SST24 J143025.7+342957	& 2.545  & 2.471 & 15.23 & 4 (LGS)& 69 & 0.04& 19.3& 0.104     & 14.9 & 30\\
DOG 13&	 SST24 J142538.1+351856	& 2.260  & 0.846 & 17.60 & 5 (LGS)& 116& 0.04& 19.5& 0.070    & 14.3 & 49\\
DOG 14&	 SST24 J143424.5+334542	& 2.263  & 0.861 & 15.56 & 5 (LGS)& 36 & 0.04& 19.0& 0.078     & 13.8 & 34\\
DOG 15&	 MIPS 16113		& 1.930  & 1.042 & 14.20 & 3 (NGS)& 66 & 0.04& 19.1& 0.075 &     12.9 & 31\\
%DOG 16&  MIPS 8184		& 0.990  & 1.540 & 13.31 & 3 (NGS)& 21 & 0.04 & 0.075 &    & 14.5 & 20\\
\enddata			 
\tablenotetext{a}{\citet{Houck05}}
\end{deluxetable*}

Observations were made with the NIRC2 infrared camera in the $K'$ filter.  Details of the observations including exposure times, pixel scale, and estimated AO performance (full-width-half-max of the point-spread-function) are given in Table \ref{tab:obs2}.  The brightest sources were observed with the narrow ($10\arcsec\times10\arcsec$) camera, which has the finest pixel sampling ($0.01\arcsec$ pix$^{-1}$), taking full advantage of the AO correction.  Fainter DOGs, not  detected in sky subtracted 5 minute exposures with the narrow camera, were observed with the wide field camera ($0.04\arcsec$ pix$^{-1}$) which has a field of view large enough ($40\arcsec\times40\arcsec$) to contain one or more bright sources to align successive images for stacking.  

Individual images were typically between 2 and 5 minutes.  A telescope dither was applied between successive images.  Total exposure times varied from $\sim30$ minutes for the brightest sources to over an hour for the faintest sources.  One source, DOG 3, was only observed for 9 minutes because of an instrument fault.   Final reduced images were created from a clipped mean of the individual exposures after subtracting sky background, dividing flatfield variations, and correcting for camera distortions.  Sky and flatfield frames were created from the actual science images with sources masked out.  Table \ref{tab:obs2} records the per pixel surface brightness limit of each image.  Note, images taken with the narrow camera have brighter per pixel surface brightness limits, but have 16 pixels for each wide camera pixel.  

AO images of the DOGs are shown in Figures \ref{fig:im}, \ref{fig:im2}, and \ref{fig:im3}.  Contours are over-laid to show the distribution of flux more clearly.  To determine if the DOGs are resolved, the flux profiles of the DOGs are compared to AO point-spread-functions (PSFs) observed nearby in time to the target observations.  The PSFs are discussed below.

\subsection{AO Point-Spread-Function}

The AO PSF is a product of the atmosphere and the corrective optics, both of which are changing on short time scales.  Not only does the AO PSF change temporally it also changes spatially across the field.  This is  because the AO correction drops with separation from the both the tip-tilt star (anisokinetisism) and the laser spot (anisoplanatism).   Accurate understanding of the AO PSF is necessary for the morphological analysis that follows.  Therefore, in addition to observing the science targets, we also attempted to image stars, representations of the real-time AO PSF.   

For DOGs 1, 4, and 12 these PSF stars were actually in the science fields, at a similar separation from the AO guide star as the science target. For several of the NGS targets (DOGS 7, 8, 15) we observed PSF star pairs where the observed star had a similar separation from its guide star as the actual science target.  For most of the LGS targets (DOGs  2, 10, 13, 14) we observed the tip-tilt guide star as a measure of the AO PSF.  Because of anisoplanatism, observing the AO guide star is generally not a good measure of the PSF for NGS observations.  It is a more reasonable estimate for LGS observations where anisoplanatism is minimal and the isokinetic angle (the angular scale over which tip-tilt varies) is large.  For those DOGs without direct measure of a PSF (DOGs 3, 5, 6, 9, 11) we use the PSF of an alternative DOG taken nearest in time to the science observation.  Because the PSF is generally elongated in the direction of the laser spot or NGS guide star,  we rotate the PSFs to best match the expected PSF at the position of the DOG. 

The best estimate of the effective resolution of the AO images at the positions of the DOGs are given by the the full-width-half-max (FWHM) of the PSFs, presented in Table 2.  For the wide camera data, the typical FWHMs are $\sim0.08\arcsec$.  The typical narrow camera data have FWHM measures of $0.05-0.06 \arcsec$.   If the DOGs contain multiple point-like structures with separations larger than these limits they will be resolved in our images.

%\subsection{Photometric Zero Points, and $K$-band Magnitudes}
%While most of our analysis is independent of the total $K$-band flux, we do provide a measure of the $K$-band magnitude of each source.  For the photometric 2008 nights we give the $K'$ magnitude, based on standard star observations made during the observing runs.  For the non-photometric 2007 runs we base the photometry on pre-existing $Ks$-band observations of the fields.

\section{Morphology Measurements}
Figures \ref{fig:im}, \ref{fig:im2}, and \ref{fig:im3} show images of the DOGs and their associated PSFs.  A comparison of the DOG profiles with the PSF profiles suggest that the bulk of the DOGs are resolved (13 of 15).  DOGs 3 and 11, however, appear to be unresolved, or only marginally resolved.  

From these images, we make three types of morphological measurements;  (1) circular aperture photometry to estimate half light size, and concentration (defined below);  (2) 2D galaxy profile fitting with GALFIT \citep{Peng02} to measure a PSF corrected effective radius and Sersic index (useful for differentiating between disk-like and elliptical-like systems, see Section 3.2 for a definition of the Sersic profile); and  (3) multiple component fits with GALFIT to reveal any sources that show either a strong central point source indicative of AGN activity, or sources that show two or more resolved components suggestive of merging. These measurements are discussed in detail below.
\begin{figure*}
\centering
\plotone{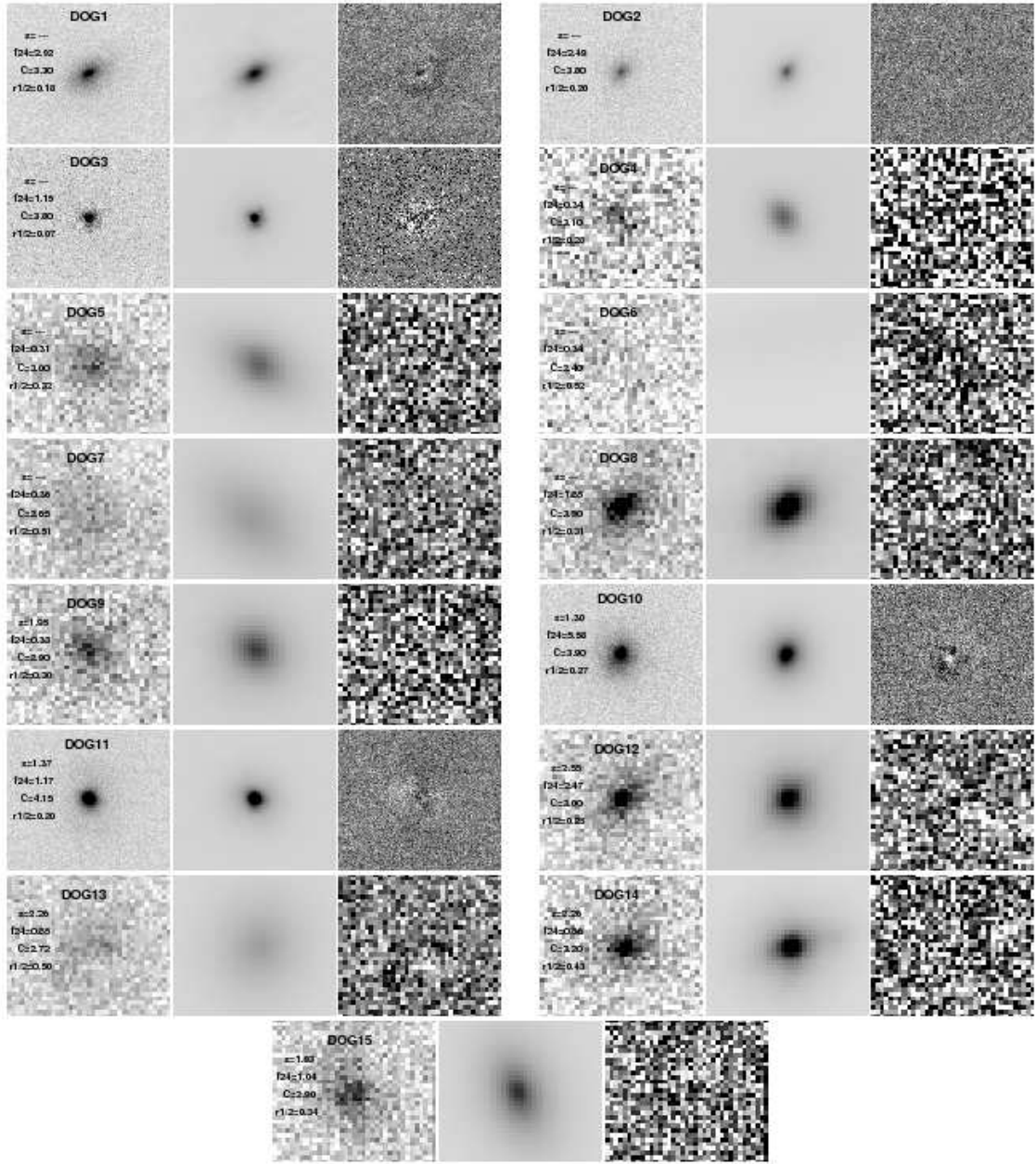}
\caption{\label{fig:images} GALFIT models for the 15 DOGs.  The left column shows the actual science data. The middle column shows the best fit single Sersic model from GALFIT \citep{Peng02}.  The right panel shows the residual difference between the two.  Images are $\sim1.5\arcsec$ on a side.}
\end{figure*}

\subsection{Circular Aperture Photometry}
For each galaxy we measure circular aperture photometry about the center of the source, where the center is selected by a Gaussian fit to the galaxy.  As was done in \citet{Bershady00} we use a curve of growth technique  to determine the total flux in each system.  We use a maximum aperture size of $2\arcsec$, 5-10 times larger than the typical DOG half-light size. From total flux, we derive the radius which contains half the light, $r_{50}$, and the galaxy concentration.  Concentration is defined as:
\begin{equation} 
	C=5log(r_{80}/r_{20}),
\end{equation} where $r_{80}$ is the radius that contains 80\% of the light, and $r_{20}$ the radius containing 20\% of the light \citep[e.g.][]{Bershady00,Conselice03}.     Concentration has been shown to correlate with galaxy Hubble type.  Typical concentrations for galaxies in the local universe range from $C\sim3$ for late type disk galaxies, to $C\sim4$ for early types and ellipticals \citep{Bershady00}.    

For the DOGs, typical concentrations range from $C\sim4$ at the high luminosity end to $C \le 3$ at the low luminosity end.  Half-light sizes vary from roughly $0.2 - 0.5 \arcsec$ or 1 to 4 Kpc (assuming the DOGs are at $z=2$). Measurements of DOG half-light radius and concentration are given in Table \ref{tab:measure}.  We do not correct $C$ and $r_{50}$ for PSF effects.  However, for comparison, we provide measurements of $C$ and $r_{50}$ of the PSF stars (Table \ref{tab:measure}).  Typical PSF concentrations are $C\sim3.9$, and half-light sizes are $r_{50}\sim0.15\arcsec$. 

Cloning local galaxies into the high redshift universe, \citet{Bershady00} showed that $C$ is robust (measured to within $10 -20$\%) for galaxies with half-light radii larger than two resolution elements (i.e. 4 pixels).  In most cases, the DOGs have half-light sizes larger than 4 resolution elements, suggesting that their $C$ measures are robust.    For the bulk of the DOGs in our sample, the  estimated uncertainty in $C$ due to small half-light sizes is significantly larger than that introduced by photometric errors.  We combine these two uncertainties as a final estimate of the uncertainty on $C$.    The uncertainties in the half-light sizes and concentrations range from 10-20\%.

\subsection{Single Sersic Profiles}

Given a galaxy image and its associated PSF, the 2D galaxy profile fitting routine GALFIT \citep{Peng02} uses a chi square minimization routine to estimate the best-fit Sersic profile of the galaxy.    The Sersic profile is defined by:
\begin{equation}
\Sigma(r)=\Sigma_e exp[-\kappa((\frac{r}{r_e})^{1/n}-1)],
\end{equation}
where $\Sigma$ is the galaxy surface brightness, $n$ is the Sersic index, and $r_e$ is the effective radius.  All profiles are assumed to be axially symmetric ellipses with $r=(x^2+y^2/q)^{1/2}$, and q equal to the minor to major axis ratio of the ellipse. A Sersic values of $n=1$ reduces to an exponential profile typically associated with disk galaxies.  A Sersic value of $n=4$ is a de Vaucouleurs profile, typically associated with elliptical galaxies.  For each DOG, GALFIT fits for total flux, Sersic index, effective radius, semi-major to semi-minor axis ratio, and position angle. GALFIT convolves each test model with the AO PSF, and minimizes the residual difference of the model image with the actual data.  Because GALFIT has trouble disentangling sky from galaxy light for the lower surface brightness edges of the galaxies, we provide GALFIT with an independent measure of the sky.  We use the median of the pixels in a $4\arcsec$ box surrounding the galaxy, with pixels associated with the galaxy removed.  
\begin{figure*}
\centering
\plotone{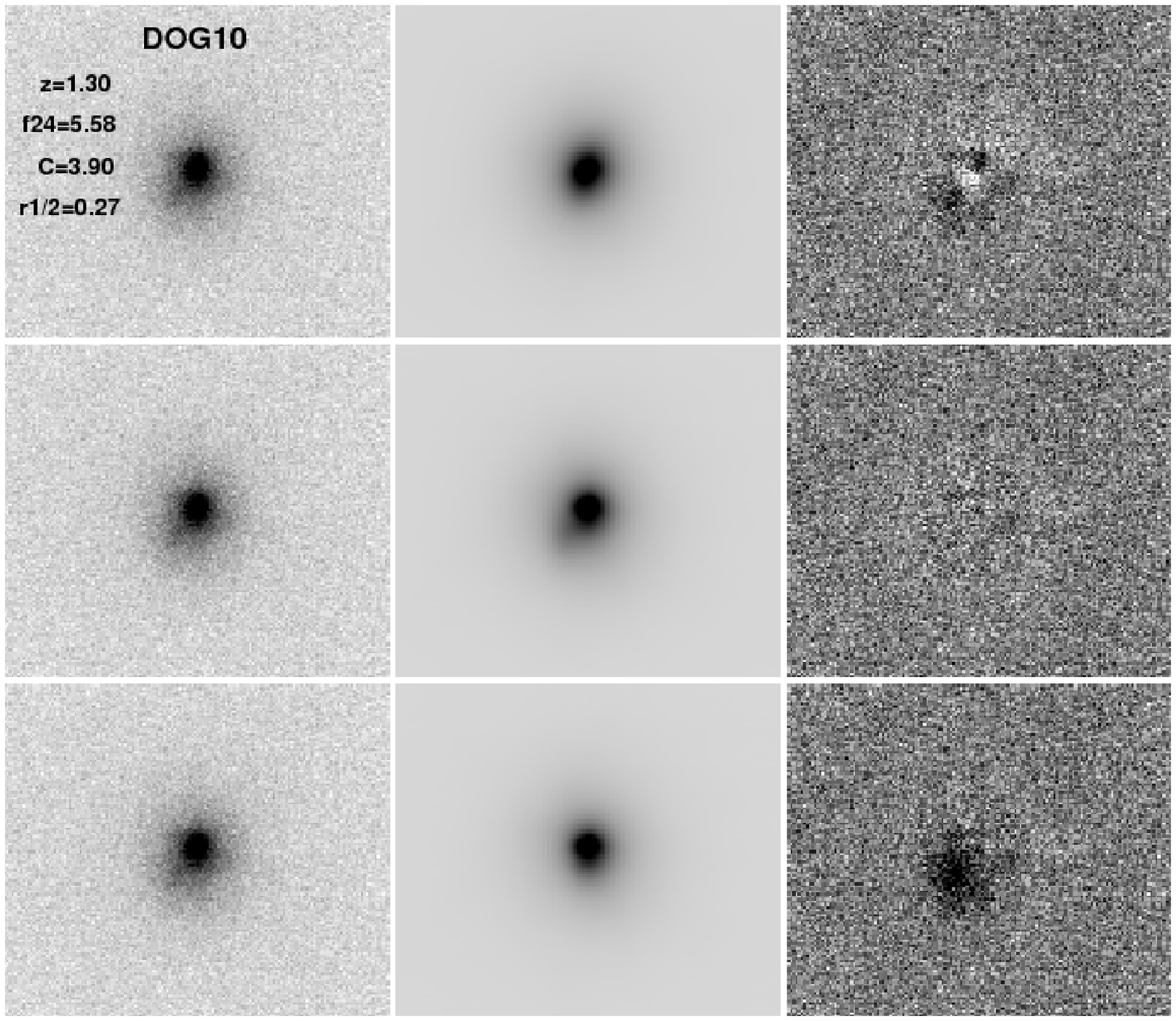}
\caption{\label{fig:merger1} Different GALFIT models of DOG 10 (middle column).  The science image is shown in the left column and the residual difference between the data and the model is shown in the right hand column.  The top row uses a single Sersic model, which is not a good fit to the galaxy light.  The middle row uses the best fit 3 component model: Sersic 1 +  PSF + Sersic 2. This model shows minimal residuals.  The bottom row uses the same 3 component model, only Sersic 2 is not subtracted off.  The two resolved components are separated by only $0.13\arcsec$ ($\sim1$ kpc at this redshift), suggesting that DOG 10 is undergoing a merger. The box size is $\sim 1.5 \arcsec$ on a side.  }
\end{figure*}

\begin{figure*}
\centering
\plotone{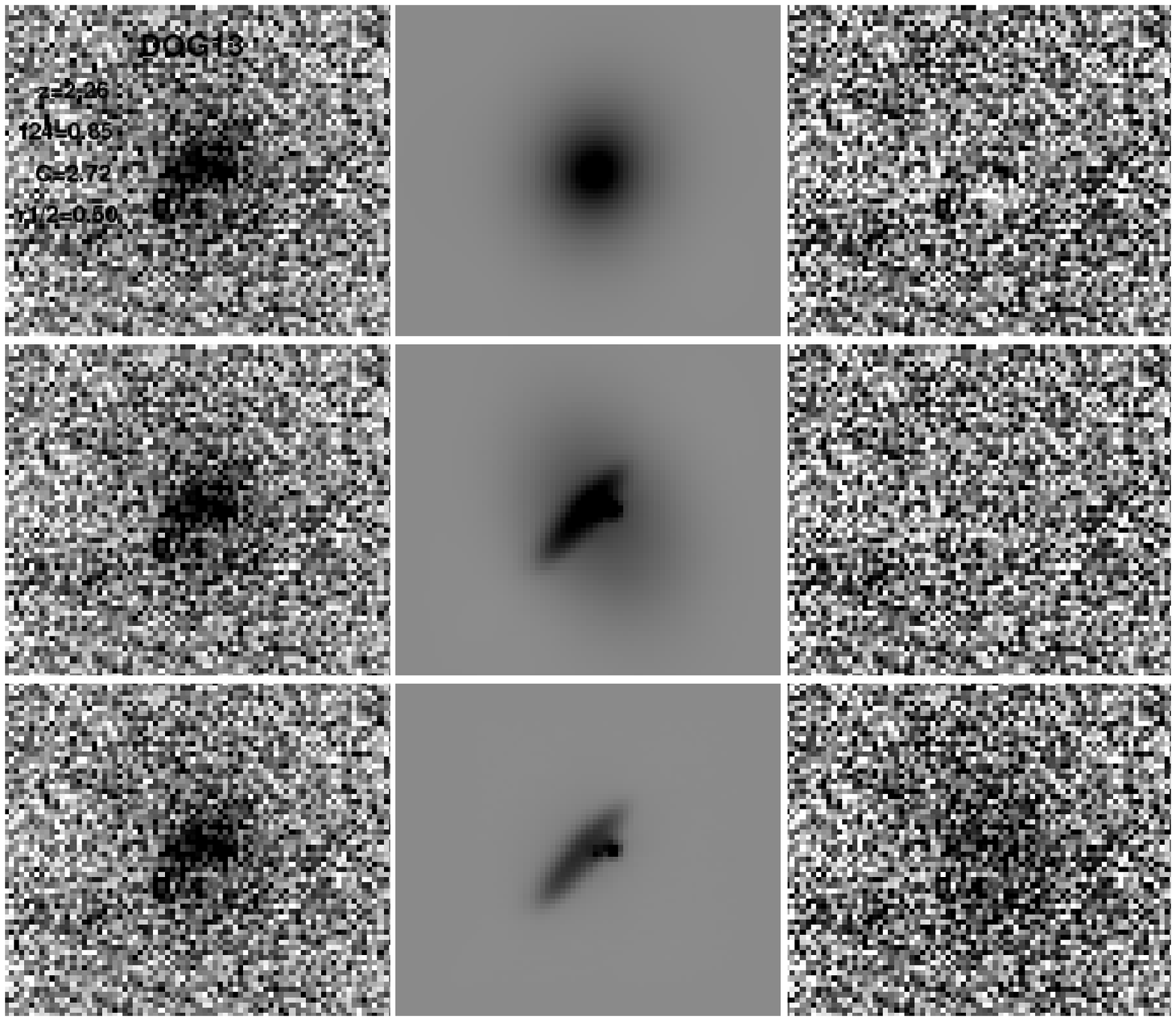}
\caption{\label{fig:merger2} Same as Figure \ref{fig:merger1} only now for DOG 13.  The single Sersic model (top row) is not a good fit.  The three component model, Sersic 1 + PSF + Sersic 2, is the best fit model (middle row).    The bottom row shows the same as the middle row, only the main galaxy is not subtracted.  This third panel is surprising.  The main galaxy is low surface brightness and not obvious until after subtracting off the companion, and yet it has 5 times the total flux of the companion.  The PSF is associated with the main galaxy. Note, this set of images is zoomed out from the previous images to show a larger area ($3\arcsec$ on side).}
\end{figure*}

The typical DOG single Sersic profile has a Sersic index of $n<2$ (9 of 14 that were measurable), although one of these systems, DOG 11, was consistent with a point source (i.e. effective radius $r_{eff} < 0.05\arcsec$). Several had a Sersic index  $n>3$ (5 of 14), although 1 of these systems, DOG 3, was also consistent with a point source.  DOG 6 was too low surface brightness to measure with GALFIT. The model parameters, Sersic index, and effective radius, are recorded in Table \ref{tab:measure}.  Because the half-light sizes of the DOGs are small,  typically only 50\% - 100\% larger than the PSF half-light sizes, the effective radii measured by GALFIT (which corrects for the PSF) are typically smaller than the DOG half-light sizes calculated with aperture photometry and no correction for the PSF.

Figure \ref{fig:images} shows the DOG images (left), the best fit single Sersic model for each DOG (middle), and the residual difference between the two (right).  For the majority of DOGs, these residuals show minimal structure.  However, several show significant structure possibly indicating a central point source or multi-component system.  Most notable of these are DOGs 1, 10, and 13.  We will examine these structures in more detail in the following sections.

While by-eye examination of the residuals is the primary method for determining the adequacy of the fits, we also calculate a quality statistic, Q, given by:

\begin{equation}
Q=\frac{\mathrm{stdev \ of \ residuals  \ within \ }0.4 \arcsec \mathrm{ \ of \ the \ galaxy \ center}}{\mathrm{stdev \ of \ pixels \  in \ the \ neighboring \ sky}}
\end{equation}

For a good fit, Q approaches 1.  This statistic was chosen because the dominant source of uncertainty was the infrared thermal sky flux.  The sky background can change dramatically on short time scales, and it was  removed during image processing.   By comparing the GALFIT residual image to noise in the sky, Q more adequately describes the quality of the fit, than the GALFIT chi-square measures which do not account for the removed sky background. For each GALFIT result, Table \ref{tab:measure} provides a Q measure.  For the single Sersic fits, typical Q values range from Q$=1$ to 2.  

\subsection{Sersic + PSF Profiles}
We also use GALFIT to estimate contribution of central point sources to the DOG light profiles.    An un-resolved point source in the center of a galaxy could indicate the presence of an AGN, unresolved bulge, or unresolved central star burst with a size less than 1 kpc. In this mode, GALFIT simultaneously fits for a Sersic + PSF profile.  Table \ref{tab:measure} gives the total flux ratio of the best fit central point source to the rest of the Sersic component.  We caution that while GALFIT produces a measurement, these two component models are not necessarily better than a one component fit.  

To interpret the results, we compare the two component fits to the single Sersic fits, looking for improvements in Q and in the residual images. For instance, DOG 10 shows a large improvement in Q when adding an additional point source component.  As will be discussed in the next section DOG 10 is actually best fit by a 3 component system Sersic + Sersic + PSF (See Figure \ref{fig:merger1}).  The point source component is roughly $\sim18$\% of the luminosity of the Sersic component.  
\begin{figure}
\centering
\plotone{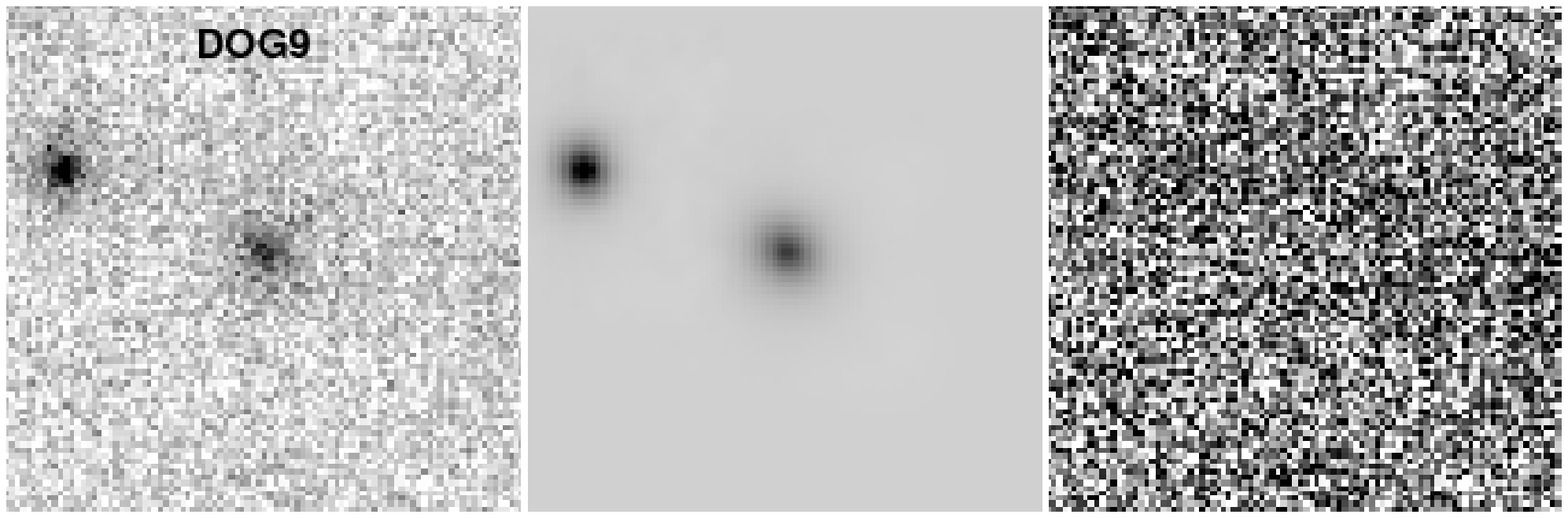}
\caption{\label{fig:dog9} A wide field view of DOG 9 reveals a second object with a projected separation of $1.6\arcsec$ ($\sim 13$ kpc if both at $z=2$).
}
\end{figure}

DOG 12 also appears to host a strong central point source.  For DOG 12 the point source has a PSF/Sersic flux ratio of 80\%.    The single Sersic fit for DOG 12 had a Sersic index of 3.47, consistent with an elliptical galaxy.  However, after including a PSF in the fit, GALFIT finds that the extended component has a Sersic index of 0.55, more consistent with a disk.  This is the only galaxy in the sample for which including a point source in the fit,  fundamentally changes the Sersic index from elliptical-like to disk-like.

Four other DOGs have PSF/Sersic flux ratios greater than 20\%, (DOGs 3, 4,  8, and 11).  However,  in the case of DOG 4, which has the largest PSF/Sersic flux ratio of 2.6, the Q value and the residual image is actually worse for the 2 component model than for the 1 component model.  Therefore, we do not believe that the two component, PSF dominated, model is an accurate representation of the galaxy.  

For DOGs 3 and 11, the single Sersic fits were already consistent with point sources.  Recall, the single component Sersic fits to DOGs 3 and 11 had effective radii of $0.02\arcsec$ and $0.03\arcsec$ respectively, i.e. the size of PSFs.   Therefore, any additional components were difficult to measure.  When GALFIT fit for an additional component in the DOG 3 model, it measured a nonsensical size of 90 arcsec for it (see Table \ref{tab:measure}); usually this indicates that GALFIT has confused sky with object.  Therefore the single component, point source dominated fit is a better approximation for the morphology of DOG 3 than the two component model.  In the case of DOG 11, the GALFIT two component Sersic + PSF model has a Sersic component with an effective radius the size of a PSF.   Thus GALFIT effectively chose to fit the galaxy with a  PSF + PSF model.  This again suggests that the single Sersic fit is more appropriate than a two component fit. Thus, both DOGs 3 and 11 appear to be point-like and any underlying resolved galaxy component is too faint to detect.

In summary, five DOGs show strong evidence for luminous, central, point-like structures (DOGs 3, 8, 10, 11, and 12).  GALFIT finds less convincing evidence for luminous point-like structures in  the remaining 9 systems that were measurable.
 \begin{deluxetable*}{c|cc|ccc|ccc|cccc|cccc}
\tabletypesize{\tiny}
\tablecaption{Morphology Summary \label{tab:measure}}
\tablehead{\colhead{object} & \multicolumn{2}{c}{PSF} & \multicolumn{3}{c}{Galaxy}&\multicolumn{3}{c}{Single Sersic Model} & \multicolumn{4}{c}{Sersic $+$ PSF Model} & \multicolumn{4}{c}{Multiple Sersic Model}\\ 
&  \colhead{C \tablenotemark{a}}&\colhead{$r_{50}$ } &   \colhead{C \tablenotemark{a}}& \colhead{$r_{50}$ } & \colhead{$R_{1/2}$ }  &\colhead{n\tablenotemark{b}} & \colhead{$r_{eff}$ } &\colhead{Q\tablenotemark{d}}&\colhead{n} &\colhead {$r_{eff}$ } &\colhead{flux ratio\tablenotemark{e}} &\colhead{Q\tablenotemark{d}} &\colhead{flux ratio \tablenotemark{f}}&\colhead {Sep }&\colhead {Sep }&\colhead{Q\tablenotemark{d}}\\ 
&& \colhead{[$\arcsec$]} &&\colhead{[$\arcsec$]} & \colhead{[kpc]}& &  \colhead{[$\arcsec$]} & &&\colhead{[$\arcsec$]}&&&&\colhead{[$\arcsec$]}&\colhead{[kpc]}}

%\tablehead{\colhead{object} & \colhead{z} &\colhead{$f_{24}$ [mJ]} & \colhead{C \tablenotemark{1}}  & \colhead{$R_{1/2}$ [$\arcsec$] }& \multicolumn{2}{c}{Single Sersic Model}}
\startdata
DOG 1 & 	4.3& 0.10& 3.3&0.18 &1.51 \tablenotemark{c}&	1.54&	0.11&	1.72&       1.50&	0.11& 2.44E-03   & 1.71 &   6.66E-01 & -&-& 1.26 \\
DOG 2 & 	5.0& 0.10& 3.8&0.20 &1.67 \tablenotemark{c}&	3.48&	0.09&	1.03&       3.47&	0.09& 3.53E-02   & 1.01 & 	-	 &- &	-& -  \\
DOG 3 & 	5.0& 0.10& 3.8&0.07 &0.59 \tablenotemark{h}&	6.00&	0.02&	2.00&       8.85&      90.90\tablenotemark{g}& 3.02E-01   & 1.89 & 	-	 &- & -&-  \\
DOG 4 & 	3.9& 0.16& 3.1&0.28 &2.01 \tablenotemark{c}&	1.15&	0.08&	1.04&       0.08&	0.33& 2.61E+00  & 1.07 &	       -	         &- & -&-   \\
DOG 5 & 	3.9& 0.16& 3.0&0.32 &3.10 \tablenotemark{c}&	1.77&	0.34&	1.14&       1.16&	0.42& 1.13E-01   & 1.08 & 	-	 &- &	-&-    \\
DOG 6 & 	3.9& 0.16& 2.4&0.52 &3.40 \tablenotemark{c}&	-   &	-   &	-   &	-   & -	       & 	-	  & 	-    & 	-   &-&- \\
DOG 7 & 	3.3& 0.17& 2.6&0.51 &4.27 \tablenotemark{c}&	0.88&	0.50&	1.19&       0.70&	0.53& 3.98E-02   & 1.16 &	        -	 & -&- &-   \\
DOG 8 & 	3.6& 0.13& 3.5&0.31 &2.60 \tablenotemark{c}&	4.77&	0.21&	1.44&       2.26&	0.30& 2.27E-01   & 1.38 & 	-	 &- &- &-    \\
DOG 9 & 	3.6& 0.13& 2.9&0.30 &2.55                  &	1.61&	0.21&	1.04&       0.98&	0.26& 1.24E-01   & 1.01 &	9.91E-01  &1.62 & 13.59 &- \\
DOG 10& 	4.0& 0.17& 3.9&0.27 &1.93                  &	1.78&	0.08&	1.95&       1.65&	0.08& 1.84E-01   & 1.53 &	2.21E-01  &0.13 & 1.09 & 1.28  \\
DOG 11& 	4.0& 0.17& 4.2&0.20 &1.65 \tablenotemark{h}&	0.32&	0.03&	1.61&       0.20&	0.03& 2.36E-01   & 1.55 &	1.00E-01  &0.28 & 2.36 & 1.57 \\
DOG 12&         3.6& 0.23& 3.0&0.25 &1.87                  &	3.47&	0.07&	1.16&       0.55&	0.32& 8.95E-01   & 1.11 &	-	  & 	-    & 	-  &-  \\
DOG 13&         4.2& 0.10& 2.7&0.50 &4.20                  &	0.94&	0.46&	1.13&       0.79&	0.49& 2.96E-02   & 1.10 &	2.27E-01  & 	0.16 & 	1.32 & 1.07 \\
DOG 14&         3.7& 0.23& 3.2&0.43 &3.50                  &	3.45&	0.11&	0.91&        3.52&	0.11& 4.49E-02   & 0.93 &	-	  & 	-    & 	-&-    \\
DOG 15& 	4.5& 0.13& 2.9&0.34 &2.84                  &	1.58&	0.41&	1.17&        1.40&	0.43& 3.60E-02   & 1.07 &	-	  & 	-    & 	- &-   \\
%DOG 16&0.990 & 1.540 &	2.9&	0.18 & 	1.60&	0.11&	0.85&	0.13& 1.50E-01 & 	-	  & 	-     \\
\enddata			 
\tablenotetext{a}{Concentration}
\tablenotetext{b}{Sersic Index}
\tablenotetext{c}{Assuming $z=2$}
\tablenotetext{d}{(stdev of residuals in 0.4$\arcsec$ radius aperture about the center of the galaxy) / (stdev of pixels in the neighboring sky)}
\tablenotetext{e}{(flux of point source) / (flux of Sersic)}
\tablenotetext{f}{(flux of fainter Sersic) / (flux of brighter Sersic)} 
\tablenotetext{g}{This is a very large size and probably un-physical.  This galaxy appears to be point source dominated and after subtracting off this point source GALFIT struggles to fit another component.}
\tablenotetext{h}{For the two unresolved DOGs, $R_{1/2}$ is an upper limit on the physical size.}
\end{deluxetable*}

\subsection{Multiple Resolved Components}
In the third round of GALFIT modeling, we examine each DOG for  an additional resolved component.  In Mel08, we showed that DOGs 1-3 did not show evidence for a second offset resolved component such as would be seen in an ongoing merger event.  Mel08 also showed examples of what a second resolved component would appear like in the GALFIT residual images.  In this larger data set,  DOGs 10 and 13 show residual patterns suggestive of a second resolved component.  

The best fit model for DOG 10 is actually a 3 component model: the main galaxy (Sersic 1),  an unresolved point source associated with the main galaxy (PSF), and a second resolved component (Sersic 2).  The total flux ratio of Sersic 2 to Sersic 1 is $\sim$20\%.  If any of these components are not included in the fit, there is significant structure in the residuals, and the quality code, Q, degrades.  Figure \ref{fig:merger1} shows the difference between a single Sersic fit, and a Sersic 1 + PSF + Sersic 2 fit.  Also shown is the residual without subtracting off the second resolved component.  Clearly a second resolved component is contributing to the light profile of DOG 10.  

  DOG 13 is also best fit by a 3 component model,  Sersic 1 + PSF + Sersic 2 (Figure \ref{fig:merger2}).  In this system, the component that contains the bulk of the light is low surface brightness, and not obvious in the image until the higher surface brightness component is subtracted off (to best see the low surface brightness component look at the residuals in the bottom row of Figure \ref{fig:merger2}).  It has a large half-light size, $\sim0.49 \arcsec$, and contains about 5 times the total flux of the higher surface brightness component.   The central PSF component contains only 2\% of the flux of the main galaxy.    There also appear to be several additional nearby structures not included in the fit -- e.g. a faint structure appears about $\sim0.8\arcsec$ to the south-east of the main galaxy.  It is not clear if these are additional components in the same system or just background noise. 

Several other DOGs may have companions.  
\begin{itemize} 
\item Figure \ref{fig:dog9} reveals that DOG 9 has a neighbor, with similar flux, at a separation of $1.6\arcsec$ ($\sim13$ kpc if both at $z=2$).  We do not know if this companion is at the same redshift as DOG 9.  
\item  DOG 11, which is point source dominated, shows residuals suggestive of a very faint companion 1/10 of the flux of the main system. (See Figure 5.  The possible companion is a fuzzy patch to the north of the main galaxy in Figure 5.)  Because the flux ratio is so small it is not clear if this is truly an additional system.  
\item DOG 6 is very diffuse and GALFIT was not able to fit the galaxy, but the residuals (Figure 5) after subtracting off the sky suggest that it may be made up of faint multiple components.   
\item DOG 1 also shows significant structure in the GALFIT residual image (Figure 5).  This is best accounted for by adding a second component to the fit.  This second component is consistent with an exponential bulge, rather than a merging companion.  
\end{itemize}

For the DOGs that show two resolved components Table \ref{tab:measure} gives the total flux ratio of those two components, and their separations from the main galaxy.
  
\section{Discussion}
Mel08 analyzed  DOGs 1-3, among the most luminous in the sample.  Compared with other high redshift galaxy samples, the 3 DOGs in Mel08  showed high concentrations and small half-light radii.  For example, radii were significantly smaller than the sizes of $z=1$ luminous infrared galaxies (LIRGs, $L_{IR} \sim 10^{11-12} \; L_{\odot}$) observed with Keck AO by the Center for Adaptive Optics Treasury Survey  \citep[CATS,][]{Melbourne08a}. Concentrations were higher than both the LIRGs and $z=2$ sub-mm sources \citep[e.g.][]{Pope05}.  

\begin{figure*}
\centering
\plottwo{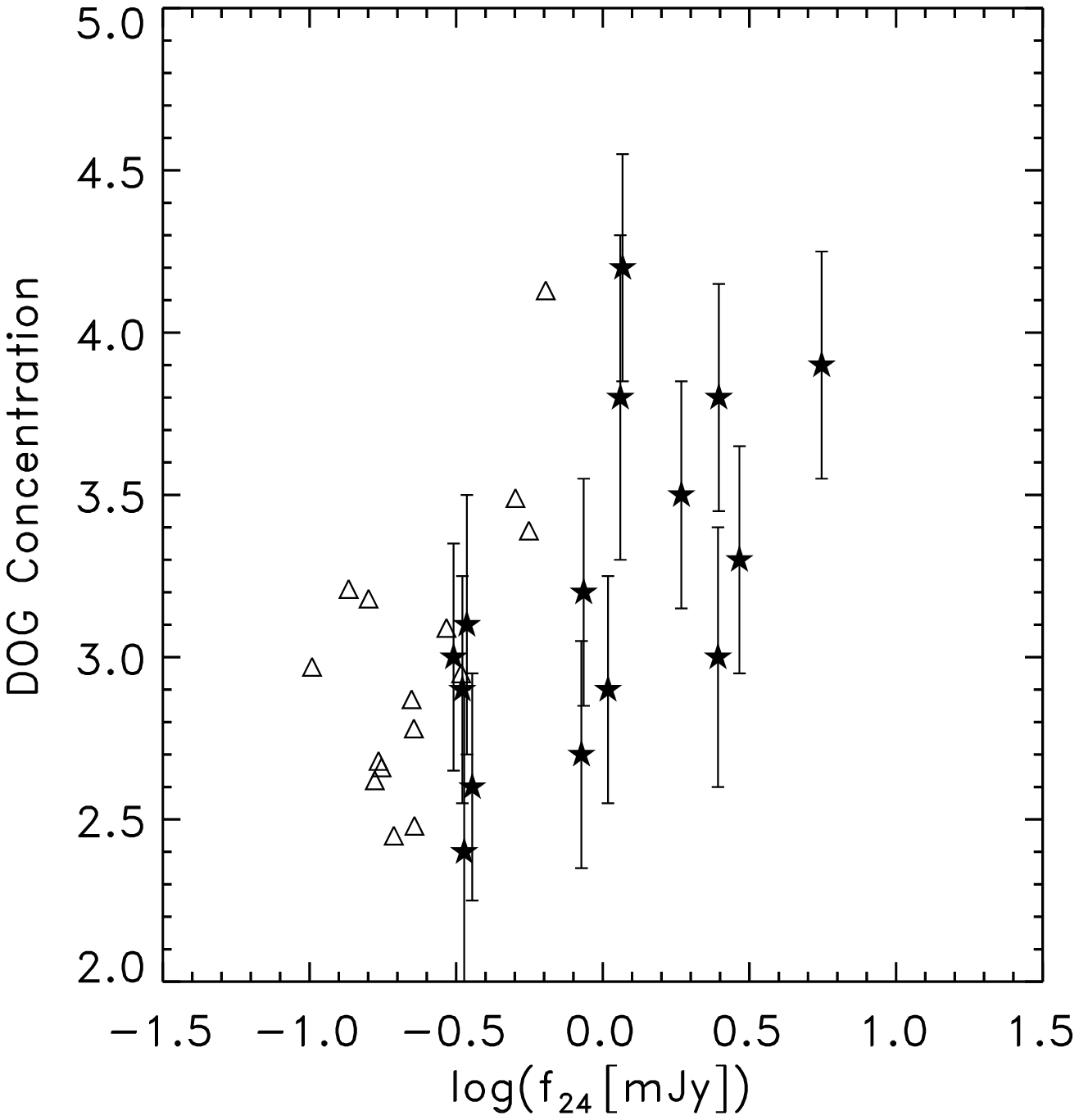}{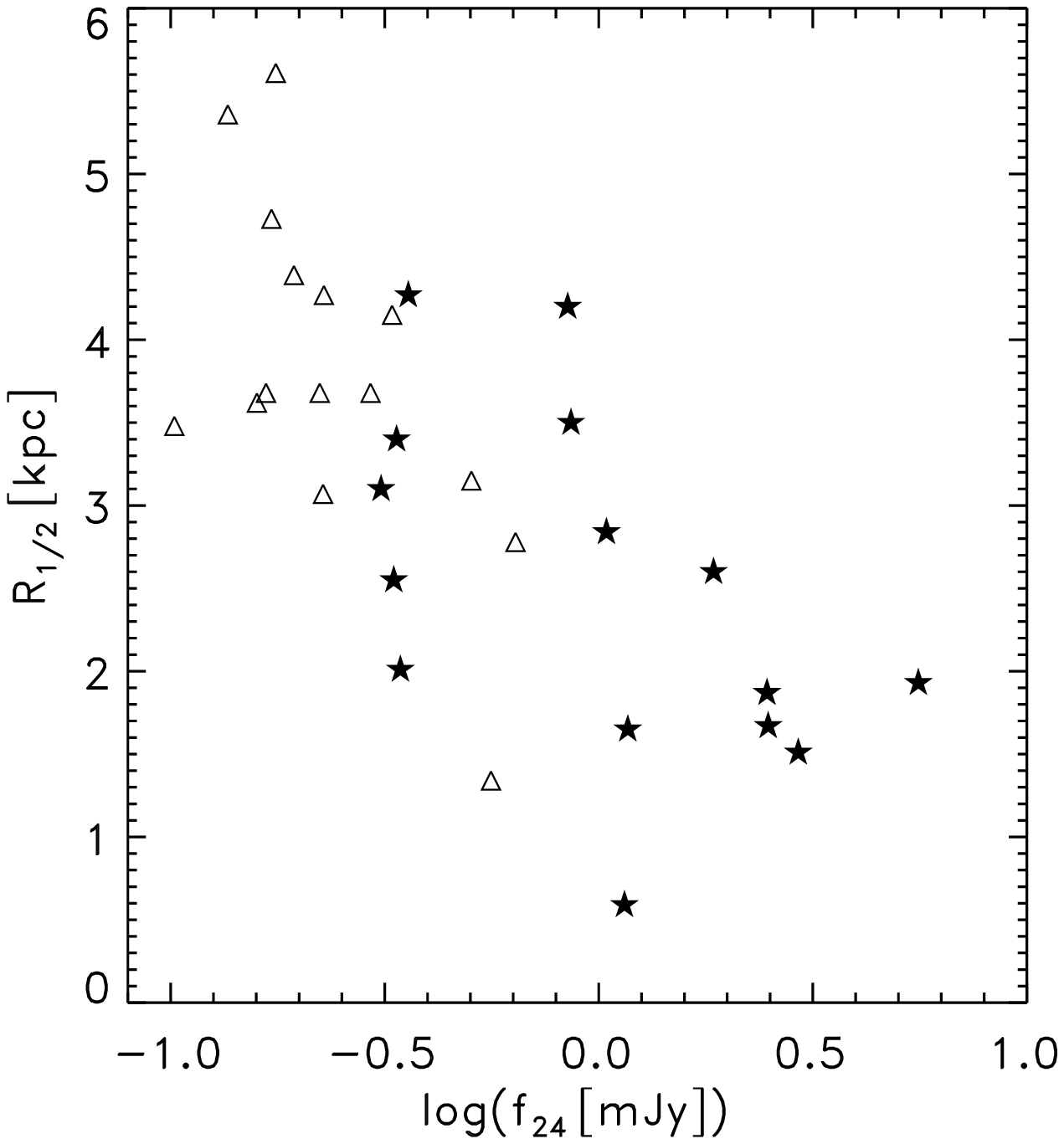}
\caption{\label{fig:compflux} Concentration (left) and half-light radius (right) as a function of \tf\ flux density. While \tf\ flux is not an intrinsic quantity, DOGs have a tight redshift distribution \citep[$<z>\sim2, \; \sigma_z\sim0.5$,][]{Dey08}, and therefore \tf\ flux should be closely tied to rest-frame mid-IR luminosity. DOGs are shown as filled stars.  A comparison sample of $z=1$ LIRGs from the the CATS survey is also shown \citep[triangles;][]{Melbourne08a}.  For the DOGs, Concentration appears correlated with \tf\ flux.   A chi square test rules out, at the 99\% level, a constant fit (i.e. no correlation) to the C vs. \tf\ data.   A similar trend is seen for half-light size, with the more luminous systems tending to show smaller sizes.  The more luminous DOGs are typically smaller with higher concentrations than the $z=1$ LIRGs.}
\end{figure*}

\begin{figure*}
\centering
\plottwo{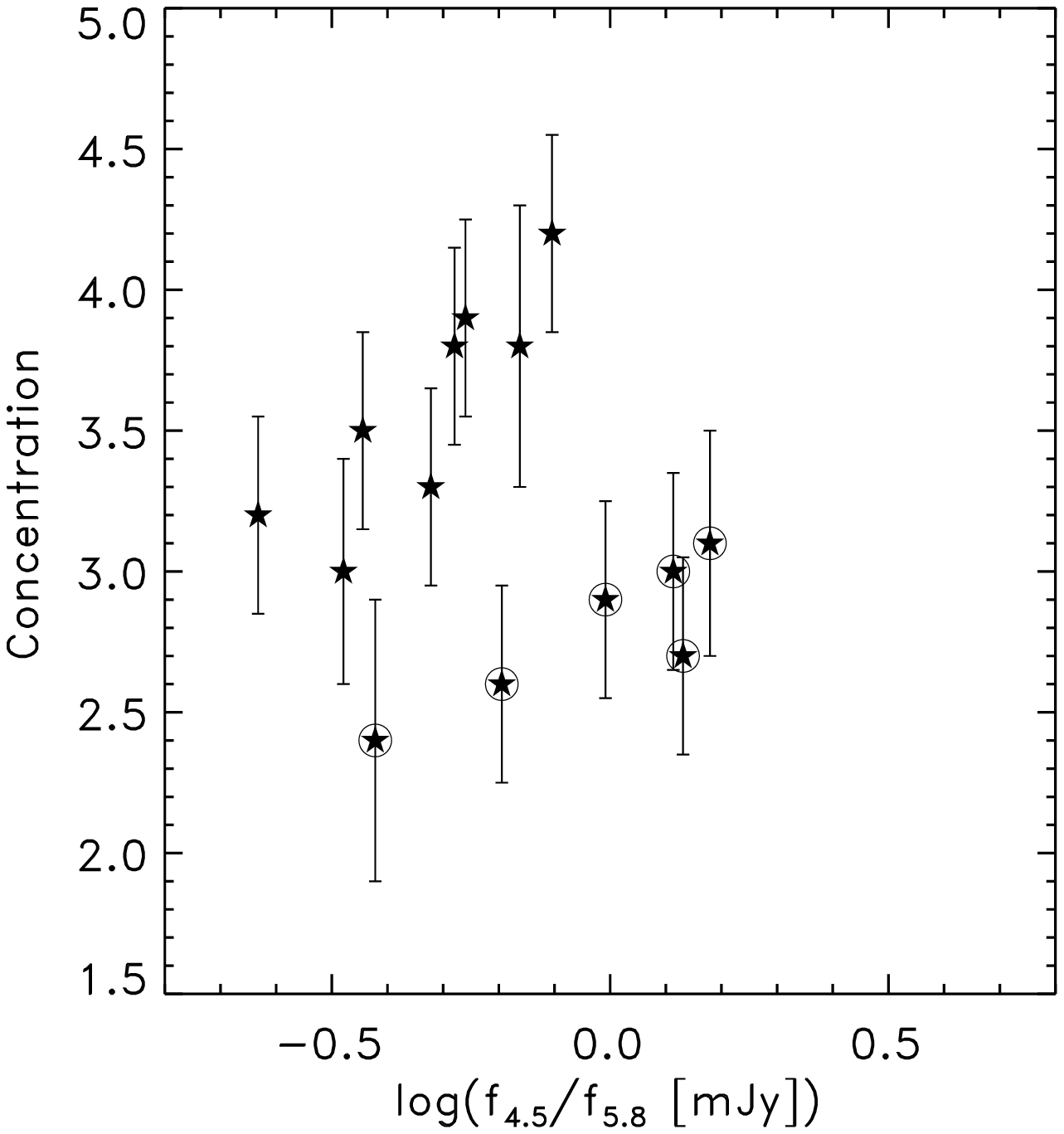}{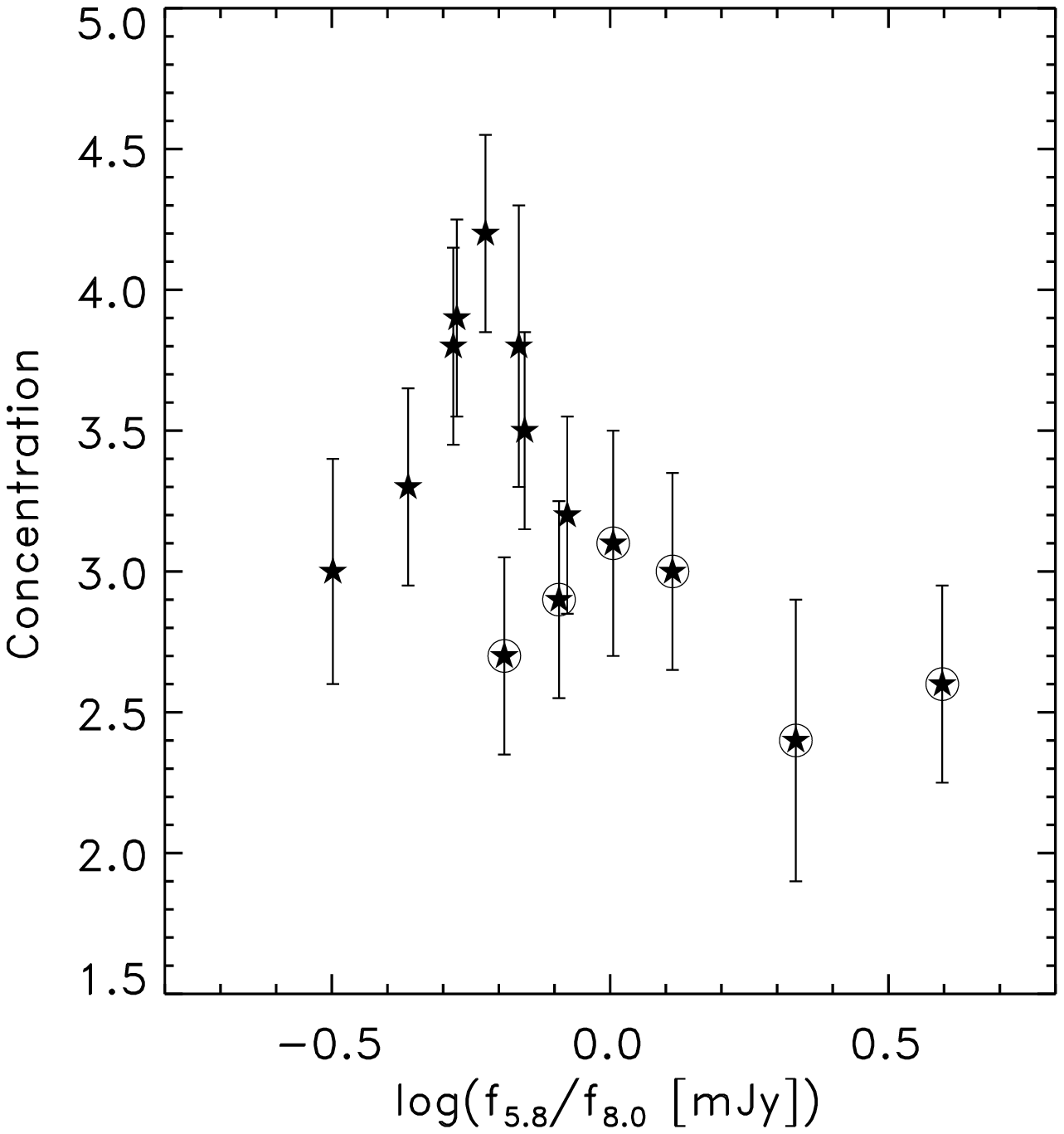}
\caption{\label{fig:compcolor} Concentration vs. IR color \citep[from \Spitzer\ IRAC imaging;][]{Eisenhardt04}.  $f_{\nu}$(4.5 \um) / $f_{\nu}$(5.8 \um)  is shown on the left, while $f_{\nu}$(5.8 \um) / $f_{\nu}$(8.0 \um)  is shown on the right.  ``Bump'' sources, with a rest 1.6 \um\ SED stellar bump  are likely to have a log color ratio greater than 0 in at least one of these two plots (the six bump candidates are circled).  Power-law sources, more likely to be powered by AGN, should show log color values less than 0 in both these plots.  Interestingly, the most diffuse systems appear to have color terms suggestive of SED bumps. The two most diffuse sources appear to show a bump at 5.8 \um (right).   The potentially 4 additional bump sources at 4.5 \um\ (two of which appear bump like at 5.8 \um\ as well),  also exhibit diffuse morphologies (left).}
\end{figure*}

In this work, the significantly larger sample spans the full range of 24 micron flux densities of DOGs in \boot. Although it contains a higher percentage of DOGs at the high flux end, the sample includes DOGs that are almost order of magnitude fainter than the DOGs in Mel08.  While many of the brighter DOGs show high concentrations (similar to Mel08), as evidenced by radial profiles only marginally more extended than PSFs (Figure \ref{fig:im}), several of the fainter DOGs show substantially more diffuse morphology.

\subsection{A Morphology - Luminosity - Color Relation}
Figure \ref{fig:compflux} plots galaxy concentration (at 2.2 \um) vs. \tf\ flux density.    While flux is an observed quantity, DOGs in \boot\ have been shown to have a tight redshift distribution \citep[$<z>\sim2, \; \sigma_z\sim0.5$,][]{Dey08}, and therefore flux should be closely tied to physical luminosity.  Galaxy concentration appears to be correlated with \tf\ flux. Brighter DOGs have higher concentrations than the fainter ones.   A chi squared test, fitting concentration vs. \tf\ flux density, rules out a constant fit to the DOG data (i.e. no correlation) at the 99\% level. A linear relation between the two parameters can not be ruled out by the chi squared test.  Also shown in Figure \ref{fig:compflux} are $z\sim1$ LIRGs from the CATS sample.  The LIRGs appear to continue the trends seen in the DOG sample.

There is some concern that the statistically significant correlation between galaxy concentration and \tf\ micron flux is not physical but rather the result of the way the data were taken.  The bulk of the low \tf\ DOGs were observed in NGS mode which, at large separations from the guide star, can result in a significantly poorer AO correction compared with LGS data.  However, there is not a statistically significant correlation between PSF concentration and galaxy concentration.  Likewise a chi-square test of a fit between PSF concentration and galaxy \tf\ flux cannot rule out a constant fit (i.e. no correlation).  Therefore the correlation between galaxy concentration and \tf\ flux appears to be real and unrelated to deficiencies in the data.

Figure \ref{fig:compflux} also shows the half-light sizes (measured at 2.2 \um) of the galaxies plotted vs. \tf\ flux density.  The brightest systems tend towards smaller sizes compared with the fainter systems.  As was shown in Mel08, the brighter sources are smaller than the  $z\sim1$ LIRGs.  The fainter DOGs, however, tend to have sizes comparable to the LIRGs.

The spectral energy distributions of the DOGs may help explain these trends.  Figure \ref{fig:compcolor} shows concentration plotted as a function of  rest-frame near-IR color from Spitzer IRAC photometry.  In these plots, rest-frame near-IR color is used as a proxy for SED shape.  DOGs with power-law SED shapes (i.e. rising flux density to longer longer wavelengths) should have log total flux ratios less than 0 in both of the these plots.   However, SEDs with a strong  flux excess from cool stellar atmospheres (i.e. ``bump'' sources) should show log total flux ratios greater than 0 in one or both of these two plots.  Figure \ref{fig:compcolor} shows that the six most diffuse DOGs have a potential stellar bump. Four (DOGs 4, 5, 9, and 13) have a flux excess at 4.5 microns (left), two of which (DOGS 4 and 5) also have a flux excess at 5.8 microns (right).  The two most diffuse DOGs (DOGs 6 and 7) have a flux excess at 5.8 microns.   The other nine DOGs do not have SED shapes suggestive of bump sources.  Their SEDs rise into the IR as expected for power-law sources.

As has been discussed in previous papers, DOGs with power-law SEDs are associated with significant AGN activity \citep{Houck05,Weedman06,Brand07}.  On the other hand, star formation dominates the mid-IR flux output of DOGs with restframe near-IR SED bumps \citep{Farrah08}.    The morphologies are consistent with this picture.  DOGs with power-law SEDs, thought to be AGN dominated, tend to show higher concentrations and small physical size.   An AGN may be contributing to the centrally concentrated light in power-law DOG morphologies.  Of the five DOGs with the highest concentration (DOGs 2, 3, 8, 10, and 11), GALFIT finds evidence for a significant point  source (PSF/Sersic flux ratio greater than 18\%) in four.    DOGs with SED bumps, thought to be star formation dominated, are more diffuse.  If  AGN exist in the bump DOGs, they appear to be sufficiently enshrouded so as to not bias the rest-frame optical morphologies to higher concentrations. Of the six potential bump sources, GALFIT finds none with PSF/Sersic ratios greater than 12\%, except for DOG 4, where we suspect that the fit was poor. These results suggest a connection between the dominant power source in the system, and the morphology measured. 

Bussman et al. (2009, submitted) found similar results from \HST\  ACS, WFPC2, and NICMOS imaging.  Their paper showed that the luminous DOGs ($f_{\nu}(24) > 0.8 $ mJy) have higher concentrations than other samples of $z=2$ active galaxies including Lyman Break Galaxies and sub-mm galaxies.  The high concentration measures were more pronounced in the NICMOS $H$-band  than in the optical (rest-UV) data.  They posit that at longer wavelengths, the central AGN is able to contribute more light than at shorter wavelengths where dust obscuration is most pronounced.  The AO $K$-band images of the luminous  DOGs ($f_{\nu}(24) > 0.8 $ mJy) are even more compact than the \HST\ $H$-band ($<C_{AO}>\ = 3.4 $ vs. $<C_{HST}>\ = 3.0$), further evidence of dust obscuration at shorter wavelengths, and higher AGN contribution at longer wavelengths.

\citet{Dasyra08} also show a morphology SED connection in a set of $z=2$ ULIRGs (not necessarily DOGs).  They find a trend of galaxy size vs. 7.7 micron PAH strength.  Again PAH features are signposts of ongoing star formation.  \citet{Dasyra08} found that $z=2$ ULIRGs with strong PAH tend to have larger sizes than those with weak PAH.   This is similar to our result that the bump DOGs have larger sizes than the power-law DOGs.

\subsection{Mergers as Triggers}
By analogy with local ULIRGs, the images may reveal whether the DOGs are triggered by mergers.  Mel08 demonstrated the wide range of DOG morphologies.  The 3 systems discussed in Mel08 each exhibited different morphologies, an exponential disk, a de Vaucouleurs profile (suggestive of an elliptical galaxy), and an unresolved source.  No obvious evidence of an ongoing merger (e.g. tidal tails, or double nuclei)  was seen.

The majority of DOGs in the current sample, 8 of 14 that were fit by GALFIT, have resolved single Sersic profiles with Sersic indices $n< 2$, suggestive of disk galaxies. In the Bussman et al. (2009, submitted) \HST\ sample, the fraction of disk-like systems was even higher (27 of 29).  Only 4 of 14 DOGs in the AO sample have resolved profiles with indices, $n>2$, suggestive of elliptical galaxies, and one of those is better fit with a disk+PSF rather than an elliptical profile.  Of the remaining three DOGs in the sample, two are consistent with being point source dominated, and one was too low S/N to fit with GALFIT.    

In contrast with Mel08, this sample contains two examples of systems with multiple resolved components, circumstantial evidence for merging.  DOGs 10 and 13 show two resolved components with separations on the order of 1 kpc and total flux ratios on the order of one to five (possibly minor mergers).  They also both appear to contain an additional unresolved component, suggestive of an AGN.  DOG 10 has a power-law SED and a very high concentration.  DOG 13 has a bump-like SED and is very diffuse.  These systems suggest that mergers may trigger  both AGN and star formation dominated DOGs.  However, the majority of DOGs in the sample do not show evidence for ongoing mergers (i.e. multiple resolved components).  This conclusion was also reached in Busmann et al. (2009, submitted), which had only 5 of 30 showing evidence for an ongoing merger.  

As Bussman et al. (2009, submitted) point out, the DOG merger fraction is much lower than in the local ULIRG sample which contains 35\% with double nuclei at separations larger than 2.3 kpc (easily resolvable in both the \HST\ and AO samples).  The DOG merger fraction is also much lower than the merger fraction in the general $z=2$ ULIRG population, in which \citet{Dasyra08} found evidence for interactions in up 50\%.   Thus, if mergers are required to trigger DOGs, DOGs may preferentially exhibit the DOG photometric selection criteria only after the merger has occurred.  Alternatively, the very red optical to IR color selection may just last longer than than the morphological evidence of the merger.  It is also possible that mergers are unnecessary for the production of DOGs.

\subsection{Summary}

The morphologies of the DOGs presented in this paper appear to be correlated with their photometric properties.  As a result, a consistent picture is emerging for the nature of these extreme systems.  The most luminous, tend to show power-law SEDs and small, highly concentrated morphologies.  Both of these trends can be explained by strong AGN activity.  The less luminous systems tend to show SED bumps in the IRAC bands.  They also tend to be larger and more diffuse than the brighter systems.  The trends seen in the bump sources can be explained by a star formation dominated power source.  Mergers may trigger both the AGN-dominated sources and the bump-dominated sources, although it is not clear if mergers are necessary to produce DOGs.  

\acknowledgments
The adaptive optics data were obtained at the Keck Observatory, which is operated as a scientific partnership among Caltech, UC, and NASA.  The authors wish to recognize and acknowledge the very significant cultural role and reverence that the summit of Mauna Kea has always had within the indigenous Hawaiian community. The laser guide star adaptive optics system was funded by the W. M. Keck Foundation.     
This work is based in part on observations made with the \Spitzer\ Space Telescope, which 
is operated by the Jet Propulsion Laboratory and the California Institute of Technology under a NASA  contract.  Support for this work was provided by NASA through awards issued by JPL/Caltech. We are especially grateful to the IRAC Shallow Survey team (P.I.  Peter Eisenhardt) for providing observations in the Bo\"otes field.  We are grateful to the expert assistance of the staff of Kitt Peak National  Observatory where the ground-based Bo\"otes field observations of the NDWFS were obtained. The authors thank NOAO for supporting the NOAO Deep Wide-Field Survey. The research activities of AD  and BTJ are supported by NOAO, which is operated by the Association of Universities for  Research in Astronomy (AURA) under a cooperative agreement with the National Science Foundation.

\bibliographystyle{apj}
\bibliography{/Users/jmel/bib/bigbib2}

\begin{thebibliography}{25}
\expandafter\ifx\csname natexlab\endcsname\relax\def\natexlab#1{#1}\fi

\bibitem[{{Armus} {et~al.}(2007){Armus}, {Charmandaris}, {Bernard-Salas},
  {Spoon}, {Marshall}, {Higdon}, {Desai}, {Teplitz}, {Hao}, {Devost}, {Brandl},
  {Wu}, {Sloan}, {Soifer}, {Houck}, \& {Herter}}]{Armus07}
{Armus}, L., {Charmandaris}, V., {Bernard-Salas}, J., {Spoon}, H.~W.~W.,
  {Marshall}, J.~A., {Higdon}, S.~J.~U., {Desai}, V., {Teplitz}, H.~I., {Hao},
  L., {Devost}, D., {Brandl}, B.~R., {Wu}, Y., {Sloan}, G.~C., {Soifer}, B.~T.,
  {Houck}, J.~R., \& {Herter}, T.~L. 2007, \apj, 656, 148

\bibitem[{{Bershady} {et~al.}(2000){Bershady}, {Jangren}, \&
  {Conselice}}]{Bershady00}
{Bershady}, M.~A., {Jangren}, A., \& {Conselice}, C.~J. 2000, \aj, 119, 2645

\bibitem[{{Brand} {et~al.}(2007){Brand}, {Dey}, {Desai}, {Soifer}, {Bian},
  {Armus}, {Brown}, {Le Floc'h}, {Higdon}, {Houck}, {Jannuzi}, \&
  {Weedman}}]{Brand07}
{Brand}, K., {Dey}, A., {Desai}, V., {Soifer}, B.~T., {Bian}, C., {Armus}, L.,
  {Brown}, M.~J.~I., {Le Floc'h}, E., {Higdon}, S.~J., {Houck}, J.~R.,
  {Jannuzi}, B.~T., \& {Weedman}, D.~W. 2007, \apj, 663, 204

\bibitem[{{Conselice}(2003)}]{Conselice03}
{Conselice}, C.~J. 2003, \apjs, 147, 1

\bibitem[{{Dasyra} {et~al.}(2008){Dasyra}, {Yan}, {Helou}, {Surace}, {Sajina},
  \& {Colbert}}]{Dasyra08}
{Dasyra}, K.~M., {Yan}, L., {Helou}, G., {Surace}, J., {Sajina}, A., \&
  {Colbert}, J. 2008, \apj, 680, 232

\bibitem[{{Dey} {et~al.}(2008){Dey}, {Soifer}, {Desai}, {Brand}, {Le Floc'h},
  {Brown}, {Jannuzi}, {Armus}, {Bussmann}, {Brodwin}, {Bian}, {Eisenhardt},
  {Higdon}, {Weedman}, \& {Willner}}]{Dey08}
{Dey}, A., {Soifer}, B.~T., {Desai}, V., {Brand}, K., {Le Floc'h}, E., {Brown},
  M.~J.~I., {Jannuzi}, B.~T., {Armus}, L., {Bussmann}, S., {Brodwin}, M.,
  {Bian}, C., {Eisenhardt}, P., {Higdon}, S.~J., {Weedman}, D., \& {Willner},
  S.~P. 2008, \apj, 677, 943

\bibitem[{{Eggen} {et~al.}(1962){Eggen}, {Lynden-Bell}, \& {Sandage}}]{Eggen62}
{Eggen}, O.~J., {Lynden-Bell}, D., \& {Sandage}, A.~R. 1962, \apj, 136, 748

\bibitem[{{Eisenhardt} {et~al.}(2004){Eisenhardt}, {Stern}, {Brodwin}, {Fazio},
  {Rieke}, {Rieke}, {Werner}, {Wright}, {Allen}, {Arendt}, {Ashby}, {Barmby},
  {Forrest}, {Hora}, {Huang}, {Huchra}, {Pahre}, {Pipher}, {Reach}, {Smith},
  {Stauffer}, {Wang}, {Willner}, {Brown}, {Dey}, {Jannuzi}, \&
  {Tiede}}]{Eisenhardt04}
{Eisenhardt}, P.~R., {Stern}, D., {Brodwin}, M., {Fazio}, G.~G., {Rieke},
  G.~H., {Rieke}, M.~J., {Werner}, M.~W., {Wright}, E.~L., {Allen}, L.~E.,
  {Arendt}, R.~G., {Ashby}, M.~L.~N., {Barmby}, P., {Forrest}, W.~J., {Hora},
  J.~L., {Huang}, J.-S., {Huchra}, J., {Pahre}, M.~A., {Pipher}, J.~L.,
  {Reach}, W.~T., {Smith}, H.~A., {Stauffer}, J.~R., {Wang}, Z., {Willner},
  S.~P., {Brown}, M.~J.~I., {Dey}, A., {Jannuzi}, B.~T., \& {Tiede}, G.~P.
  2004, \apjs, 154, 48

\bibitem[{{Farrah} {et~al.}(2008){Farrah}, {Lonsdale}, {Weedman}, {Spoon},
  {Rowan-Robinson}, {Polletta}, {Oliver}, {Houck}, \& {Smith}}]{Farrah08}
{Farrah}, D., {Lonsdale}, C.~J., {Weedman}, D.~W., {Spoon}, H.~W.~W.,
  {Rowan-Robinson}, M., {Polletta}, M., {Oliver}, S., {Houck}, J.~R., \&
  {Smith}, H.~E. 2008, \apj, 677, 957

\bibitem[{{Fiore} {et~al.}(2008){Fiore}, {Grazian}, {Santini}, {Puccetti},
  {Brusa}, {Feruglio}, {Fontana}, {Giallongo}, {Comastri}, {Gruppioni},
  {Pozzi}, {Zamorani}, \& {Vignali}}]{Fiore08}
{Fiore}, F., {Grazian}, A., {Santini}, P., {Puccetti}, S., {Brusa}, M.,
  {Feruglio}, C., {Fontana}, A., {Giallongo}, E., {Comastri}, A., {Gruppioni},
  C., {Pozzi}, F., {Zamorani}, G., \& {Vignali}, C. 2008, \apj, 672, 94

\bibitem[{{Houck} {et~al.}(2005){Houck}, {Soifer}, {Weedman}, {Higdon},
  {Higdon}, {Herter}, {Brown}, {Dey}, {Jannuzi}, {Le Floc'h}, {Rieke}, {Armus},
  {Charmandaris}, {Brandl}, \& {Teplitz}}]{Houck05}
{Houck}, J.~R., {Soifer}, B.~T., {Weedman}, D., {Higdon}, S.~J.~U., {Higdon},
  J.~L., {Herter}, T., {Brown}, M.~J.~I., {Dey}, A., {Jannuzi}, B.~T., {Le
  Floc'h}, E., {Rieke}, M., {Armus}, L., {Charmandaris}, V., {Brandl}, B.~R.,
  \& {Teplitz}, H.~I. 2005, \apjl, 622, L105

\bibitem[{{Jannuzi} \& {Dey}(1999)}]{JannuziDey99}
{Jannuzi}, B.~T., \& {Dey}, A. 1999, in Astronomical Society of the Pacific
  Conference Series, Vol. 191, Photometric Redshifts and the Detection of High
  Redshift Galaxies, ed. R.~{Weymann}, L.~{Storrie-Lombardi}, M.~{Sawicki}, \&
  R.~{Brunner}, 111--+

\bibitem[{{Melbourne} {et~al.}(2008{\natexlab{a}}){Melbourne}, {Ammons},
  {Wright}, {Metevier}, {Steinbring}, {Max}, {Koo}, {Larkin}, \&
  {Barczys}}]{Melbourne08a}
{Melbourne}, J., {Ammons}, M., {Wright}, S.~A., {Metevier}, A., {Steinbring},
  E., {Max}, C., {Koo}, D.~C., {Larkin}, J.~E., \& {Barczys}, M.
  2008{\natexlab{a}}, \aj, 135, 1207

\bibitem[{{Melbourne} {et~al.}(2008{\natexlab{b}}){Melbourne}, {Desai},
  {Armus}, {Dey}, {Brand}, {Thompson}, {Soifer}, {Matthews}, {Jannuzi}, \&
  {Houck}}]{Melbourne08b}
{Melbourne}, J., {Desai}, V., {Armus}, L., {Dey}, A., {Brand}, K., {Thompson},
  D., {Soifer}, B.~T., {Matthews}, K., {Jannuzi}, B.~T., \& {Houck}, J.~R.
  2008{\natexlab{b}}, \aj, 136, 1110

\bibitem[{{Peng} {et~al.}(2002){Peng}, {Ho}, {Impey}, \& {Rix}}]{Peng02}
{Peng}, C.~Y., {Ho}, L.~C., {Impey}, C.~D., \& {Rix}, H.-W. 2002, \aj, 124, 266

\bibitem[{{Pope} {et~al.}(2005){Pope}, {Borys}, {Scott}, {Conselice},
  {Dickinson}, \& {Mobasher}}]{Pope05}
{Pope}, A., {Borys}, C., {Scott}, D., {Conselice}, C., {Dickinson}, M., \&
  {Mobasher}, B. 2005, \mnras, 358, 149

\bibitem[{{Pope} {et~al.}(2008){Pope}, {Bussmann}, {Dey}, {Meger}, {Alexander},
  {Brodwin}, {Chary}, {Dickinson}, {Frayer}, {Greve}, {Huynh}, {Lin},
  {Morrison}, {Scott}, \& {Yan}}]{Pope08}
{Pope}, A., {Bussmann}, R.~S., {Dey}, A., {Meger}, N., {Alexander}, D.~M.,
  {Brodwin}, M., {Chary}, R.-R., {Dickinson}, M.~E., {Frayer}, D.~T., {Greve},
  T.~R., {Huynh}, M., {Lin}, L., {Morrison}, G., {Scott}, D., \& {Yan}, C.-H.
  2008, ArXiv e-prints

\bibitem[{{Sanders} {et~al.}(1988){Sanders}, {Soifer}, {Elias}, {Madore},
  {Matthews}, {Neugebauer}, \& {Scoville}}]{Sanders88}
{Sanders}, D.~B., {Soifer}, B.~T., {Elias}, J.~H., {Madore}, B.~F., {Matthews},
  K., {Neugebauer}, G., \& {Scoville}, N.~Z. 1988, \apj, 325, 74

\bibitem[{{Somerville} {et~al.}(2001){Somerville}, {Primack}, \&
  {Faber}}]{Somerville01}
{Somerville}, R.~S., {Primack}, J.~R., \& {Faber}, S.~M. 2001, \mnras, 320, 504

\bibitem[{{van Dam} {et~al.}(2004){van Dam}, {Le Mignant}, \&
  {Macintosh}}]{vanDam04}
{van Dam}, M.~A., {Le Mignant}, D., \& {Macintosh}, B.~A. 2004, in Society of
  Photo-Optical Instrumentation Engineers (SPIE) Conference Series, Vol. 5490,
  Society of Photo-Optical Instrumentation Engineers (SPIE) Conference Series,
  ed. D.~{Bonaccini Calia}, B.~L. {Ellerbroek}, \& R.~{Ragazzoni}, 174--183

\bibitem[{{Weedman} {et~al.}(2006){Weedman}, {Le Floc'h}, {Higdon}, {Higdon},
  \& {Houck}}]{Weedman06}
{Weedman}, D.~W., {Le Floc'h}, E., {Higdon}, S.~J.~U., {Higdon}, J.~L., \&
  {Houck}, J.~R. 2006, \apj, 638, 613

\bibitem[{{Wizinowich} {et~al.}(2000){Wizinowich}, {Acton}, {Shelton},
  {Stomski}, {Gathright}, {Ho}, {Lupton}, {Tsubota}, {Lai}, {Max}, {Brase},
  {An}, {Avicola}, {Olivier}, {Gavel}, {Macintosh}, {Ghez}, \&
  {Larkin}}]{Wizinowich00}
{Wizinowich}, P., {Acton}, D.~S., {Shelton}, C., {Stomski}, P., {Gathright},
  J., {Ho}, K., {Lupton}, W., {Tsubota}, K., {Lai}, O., {Max}, C., {Brase}, J.,
  {An}, J., {Avicola}, K., {Olivier}, S., {Gavel}, D., {Macintosh}, B., {Ghez},
  A., \& {Larkin}, J. 2000, \pasp, 112, 315

\bibitem[{{Wizinowich} {et~al.}(2006){Wizinowich}, {Le Mignant}, {Bouchez},
  {Campbell}, {Chin}, {Contos}, {van Dam}, {Hartman}, {Johansson}, {Lafon},
  {Lewis}, {Stomski}, {Summers}, {Brown}, {Danforth}, {Max}, \&
  {Pennington}}]{Wizinowich06}
{Wizinowich}, P.~L., {Le Mignant}, D., {Bouchez}, A.~H., {Campbell}, R.~D.,
  {Chin}, J.~C.~Y., {Contos}, A.~R., {van Dam}, M.~A., {Hartman}, S.~K.,
  {Johansson}, E.~M., {Lafon}, R.~E., {Lewis}, H., {Stomski}, P.~J., {Summers},
  D.~M., {Brown}, C.~G., {Danforth}, P.~M., {Max}, C.~E., \& {Pennington},
  D.~M. 2006, \pasp, 118, 297

\bibitem[{{Yan} {et~al.}(2007){Yan}, {Sajina}, {Fadda}, {Choi}, {Armus},
  {Helou}, {Teplitz}, {Frayer}, \& {Surace}}]{Yan07}
{Yan}, L., {Sajina}, A., {Fadda}, D., {Choi}, P., {Armus}, L., {Helou}, G.,
  {Teplitz}, H., {Frayer}, D., \& {Surace}, J. 2007, \apj, 658, 778

\bibitem[{{Zirm} {et~al.}(2007){Zirm}, {van der Wel}, {Franx}, {Labb{\'e}},
  {Trujillo}, {van Dokkum}, {Toft}, {Daddi}, {Rudnick}, {Rix},
  {R{\"o}ttgering}, \& {van der Werf}}]{Zirm07}
{Zirm}, A.~W., {van der Wel}, A., {Franx}, M., {Labb{\'e}}, I., {Trujillo}, I.,
  {van Dokkum}, P., {Toft}, S., {Daddi}, E., {Rudnick}, G., {Rix}, H.-W.,
  {R{\"o}ttgering}, H.~J.~A., \& {van der Werf}, P. 2007, \apj, 656, 66

\end{thebibliography}
\clearpage

\end{document}